\documentclass[journal]{IEEEtran}
\usepackage{graphicx,amssymb,amsmath}
\usepackage{multicol}
\usepackage{CJK}
\usepackage[noadjust]{cite}
\usepackage{setspace}
\usepackage{stfloats}
\usepackage{midfloat}
\usepackage{flushend,cuted}
\usepackage{bm}
\usepackage{multirow}
\usepackage{array}
\usepackage{stfloats}
 \usepackage{color}
\usepackage{cite}
\usepackage{algorithm}
\usepackage{algpseudocode}
\usepackage{psfrag}
\usepackage{ccaption}

\IEEEoverridecommandlockouts

\ifCLASSINFOpdf
\else
\fi
\newcommand{\tr}{\textmd{tr}}
\newcommand{\E}{\textmd{E}}
\newcommand{\Var}{\textmd{Var}}
\newcommand{\diag}{\textmd{diag}}
\newcommand{\vect}{\textmd{vec}}
\newcounter{MYtempeqncnt}

\allowdisplaybreaks
\begin{document}
\title{Channel Estimation and Performance Analysis of\\ One-Bit Massive MIMO Systems}


\author{Yongzhi Li,
        Cheng Tao,~\IEEEmembership{Member, IEEE,}
        Gonzalo Seco-Granados,~\IEEEmembership{Senior Member, IEEE,}    \\
        Amine Mezghani,
        A. Lee Swindlehurst,~\IEEEmembership{Fellow,~IEEE}
        ~and~Liu Liu,~\IEEEmembership{Member, IEEE,}
\thanks{The research was supported in part by Beijing Nova Programme (Grant No. xx2016023), the NSFC project under grant No.61471027, the Research Fund of National Mobile Communications Research Laboratory, Southeast University (No.2014D05, No. 2017D01), and Beijing Natural Science Foundation project under grant No.4152043. A.~Swindlehurst was supported by the National Science Foundation under Grant ECCS-1547155, and by the Technische Universit\"at M\"unchen Institute for Advanced Study, funded by the German Excellence Initiative and the European Union Seventh Framework Programme under grant agreement No. 291763, and by the European Union under the Marie Curie COFUND Program.}
\thanks{Y. Li, C. Tao, and L. Liu are with the Institute of Broadband
Wireless Mobile Communications, Beijing Jiaotong University, Beijing, 100044, China (email:
liyongzhi@bjtu.edu.cn; chtao@bjtu.edu.cn; liuliu@bjtu.edu.cn).}
\thanks{G. Seco-Granados is with the Telecommunications and Systems
Engineering Department, Universitat Aut{\`o}noma  de Barcelona, Barcelona
08193, Spain (e-mail: gonzalo.seco@uab.es).}
\thanks{A. Mezghani and A. Swindlehurst are with the Center for Pervasive Communications and
Computing, University of California, Irvine, CA 92697 USA (e-mail: amezghan@uci.edu; swindle@uci.edu). A. Swindlehurst is also a Hans Fischer Senior Fellow of the Institute for Advanced Study at the Technical University of Munich.}
\thanks{Corresponding authors: chtao@bjtu.edu.cn, liuliu@bjtu.edu.cn.}
}

\maketitle

\begin{abstract}
This paper considers channel estimation and system performance for the uplink of a single-cell massive multiple-input multiple-output (MIMO) system. Each receive antenna of the base station (BS) is assumed to be equipped with a pair of one-bit analog-to-digital converters (ADCs) to quantize the real and imaginary part of the received signal. We first
propose { an approach for channel estimation that is applicable for both flat and frequency-selective fading,} based on the Bussgang decomposition that reformulates the nonlinear quantizer as a linear function with identical first- and second-order statistics. The resulting channel estimator outperforms previously proposed approaches across all SNRs.
We then derive closed-form expressions for the achievable rate { in flat fading channels} assuming low SNR and a large number of users for the maximal ratio and zero forcing receivers that takes channel estimation error due to both noise and one-bit quantization into account. The closed-form expressions in turn allow us to obtain insight into important system design issues such as {\color{black} optimal resource allocation}, maximal sum spectral efficiency, overall energy efficiency, and number of antennas. Numerical results are presented to verify our analytical results and demonstrate the benefit of optimizing system performance accordingly.
\end{abstract}


%
\IEEEpeerreviewmaketitle

\section{Introduction}
Massive multiple-input multiple-output (MIMO) technology is considered to be a key component for 5G wireless communications systems, and has recently attracted considerable research interest. The main characteristic of massive MIMO is a base station (BS) array equipped with many (perhaps a hundred or more) antennas, which provides unprecedented spatial degrees of freedom for simultaneously serving multiple user terminals on the same time-frequency channel. It has been shown that, with channel state information (CSI) available at the BS, relatively simple signal processing techniques such as maximum-ratio combining (MRC) or zero-forcing (ZF) can be employed to reduce the noise and interference at the terminals, and can lead to improvements not only in spectral efficiency, but in energy efficiency as well \cite{marzetta2010noncooperative,ngo2013energy,larsson2013massive,rusek2013scaling,lu2014overview}.

In most work on massive MIMO, perfect hardware implementations with infinite resolution analog-to-digital converters (ADCs) are assumed. There has been limited prior work on the impact of non-ideal hardware on massive MIMO systems including \cite{emil2014massive, emil2015massive, emil2015impact}, which studied imperfections such as phase-drifts and additive distortion, and showed that a massive number of antennas can mitigate these effects. In terms of hardware, perhaps the most important issue at the BS for massive MIMO is the power consumption of the ADCs, which grows exponentially with the number of quantization bits \cite{walden1999analog}, and {\color{black}also grows with increased sampling rates due to wider bandwidths}. For example, commercially available ADCs with resolutions of 12 to 16 bits consume on the order of several watts \cite{texas}. For massive MIMO configurations employing large antenna arrays and many ADCs, the cost and power consumption will be prohibitive, and alternative approaches are needed.

The use of low resolution (1-3 bits) ADCs is a potential solution to this problem \cite{mezghani2008analysis,nossek2006capacity,mezghani2007modified,mezghani2012an,mezghani2012efficient,singh2009on, jianhua2015capacity,singh2009multi}. In this paper, we focus on the case of simple one-bit ADCs, which consist of a simple comparator and consume negligible power (a few milliwatts).  One-bit ADCs do not require automatic gain control and linear amplifiers, and hence the corresponding {\color{black} radio frequency (RF)} chains can be implemented with very low cost and power consumption \cite{ jianhua2015capacity, singh2009multi }. It was shown in \cite{ mezghani2008analysis } that the capacity maximizing transmit signals for one-bit ADCs operating in {\color{black} single-input single-output (SISO)} channels are discrete, unlike the infinite resolution case where a Gaussian codebook is optimal. In addition, \cite{ mezghani2008analysis } showed that MIMO capacity is not severely reduced by the coarse quantization at low signal-to-noise ratios (SNRs); in particular, the power penalty due to one-bit quantization is approximately equal to only $\pi/2$ (1.96dB) in the low SNR region \cite{ nossek2006capacity }. On the other hand, at high SNRs one-bit quantization can produce a large capacity loss \cite{jianhua2014high}, but there is reason to believe that massive MIMO systems will operate at relatively low SNRs for improved energy efficiency, exploiting array gain to overcome the resulting distortion. This will be especially true as systems move to higher (e.g., millimeter wave) frequencies. In either case, the availability of accurate BS-side CSI is
indispensable for exploiting the full potential of a massive MIMO system, and an important open question is how to reliably estimate the channel and decode the data symbols under one-bit output quantization.

Several recent papers have investigated channel estimation in massive MIMO with one-bit ADCs \cite{ning2015mixed,chiara2014massive, jacobsson2015one,jacobsson2016throughput, jianhua2014channel, juncil2015near, mollen2016performance_WSA,mollen2016one,mollen2016performance}.
A millimeter wave MIMO system with one-bit ADCs was considered in \cite{ jianhua2014channel }, which proposed a modified {\color{black} expectation-maximum (EM)} channel estimator that exploits the sparsity of such channels.
In \cite{ juncil2015near } a {\color{black} near maximum likelihood (nML)} channel estimator and detector were proposed, and the nML approach was shown to improve estimation accuracy and better support higher order constellations than the EM estimators using one-bit ADCs.  However, the channel estimators and the computed rates obtained in \cite{jianhua2014channel, juncil2015near} rely on either the maximum-likelihood algorithm or on an iterative algorithm with high complexity, and their performance is difficult to theoretically quantify. \textcolor{black}{More recently, \cite{mollen2016performance} considered a low complexity channel estimator and the corresponding achievable rate for one-bit massive MIMO systems over frequency-selective channels, using a model in which the number of channel taps goes to infinity, and the quantization noise is essentially modeled as independent, identically distributed (i.i.d.) noise.}

In this paper, we focus on channel estimation and uplink performance for massive MIMO systems with one-bit ADCs. \textcolor{black}{In contrast to \cite{mollen2016performance}, we derive more general quantization noise models that are applied separately for data detection and channel estimation. One essential and unique aspect of our derivation is that the spatial correlation between the elements of the quantizer output is taken into account,  calculated using the {\emph {arcsine law}}. Our goal is to illustrate the impact of coarsely quantized ADCs, and to give an idea of the expected performance of massive MIMO systems with one-bit ADCs compared to conventional systems that assume infinite ADC resolution. Our specific contributions are summarized below.}
\begin{itemize}
    \item We focus on use of the Bussgang decomposition \cite{bussgang1952yq} to reformulate the nonlinear quantizer operation as a statistically equivalent linear system. \textcolor{black}{Contrary to previous work, we perform a separate Bussgang decomposition for the pilot and data phase as well as for each channel realization, an approach that more accurately captures the full effect of the quantization}. We derive an algorithm that we refer to as the Bussgang Linear Minimum Mean Squared Error (BLMMSE) channel estimator { for both flat and frequency-selective channel models}. We calculate the high-SNR channel estimation error floor achieved by the proposed approach under flat fading, and show via simulation that BLMMSE outperforms previously proposed methods.
    \item We derive a lower bound { for the flat-fading case} on the theoretical rate achievable in the uplink using MRC or ZF receivers based on the BLMMSE channel estimate, and we obtain a simple but tight closed-form approximation on the uplink rate {\color{black}assuming low SNR and a large number of users} that accurately approximates our empirical observations. Similar work in \cite{li2015uplink,zhang2016spectral} relied on an additive quantization noise model \cite{orhan2015low,bai2013on} to approximate the rate, but it assumed perfect rather than estimated CSI is available at the BS, which leads to an overly optimistic assessment.
    \item Using the closed-form expression for the achievable rate, we study the power efficiency of massive MIMO with one-bit ADCs and show that similar efficiency is obtained as in conventional massive MIMO. In particular, assuming $M$ antennas, we show overall system performance remains unchanged if 1) for a fixed level of CSI accuracy (training data power independent of $M$), the transmit power of each user terminal is reduced proportionally to $1/M$, and 2) power during both training and data transmissions is reduced proportionally to $1/\sqrt{M}$.
    \item We propose an {\color{black} optimal resource allocation} scheme to maximize the sum spectral efficiency of a one-bit massive MIMO system under a total power constraint. {\color{black} Numerical results indicate that the optimal training length in one-bit systems is no longer always equal to the number of users and the proposed resource allocation scheme notably improves performance compared to the case without power allocation.}
    \item We show that to achieve similar performance, a one-bit massive MIMO system employing an MRC receiver will {\color{black} require approximately 2.2-2.3 times more antennas than a conventional system if the sum spectral efficiency for both systems is optimized by employing the optimal resource allocation scheme;} for the ZF receiver, we show that to achieve the same goal, more and more antennas are needed as average transmit power increases.
\end{itemize}
{\color{black} A preliminary version of some of these results appeared in~\cite{yongzhi2016channel}}.

The rest of this paper is organized as follows. In the next section, we present the assumed system architecture and signal model. In Section III, we propose the BLMMSE channel estimator, and then based on the BLMMSE channel estimator, in Section IV we derive a simple closed-form expression for the lower bound on the achievable rate for MRC and ZF receivers {\color{black} in the low SNR region}. Using the closed-form approximation, we then consider several system design issues related to resource allocation and the number of antennas in Section V. Simulation results are presented in Section VI and we conclude the paper in Section VII.

\emph{Notation}: The following notation is used throughout the paper. Bold uppercase (lowercase) letters denote matrices (vectors); $(.)^*$, $(.)^T$, and $(.)^H$ denote complex conjugate, transpose, and Hermitian transpose operations, respectively; $||.||$ represents the 2-norm of a vector; $\tr(.)$ represents the trace of a matrix; $\diag\{\mathbf{X}\}$ denotes a diagonal matrix containing only the diagonal entries of $\mathbf{X}$; $\otimes$ represents the Kronecker product; $\left[\mathbf{X}\right]_{ij}$ denotes the $(i,j)$th entry of $\mathbf{X}$; ${\bf{x}} \sim \mathcal{CN}\left( {{\bf{a}},{\bf{B}}} \right)$ indicates that $\bf{x}$ is a complex Gaussian vector with mean $\bf{a}$ and covariance matrix $\bf{B}$; $\E\{.\}$ and $\Var\{.\}$ denote the expected value and variance of a random variable, respectively.

\section{System Model}

\begin{figure}
  \centering
  \includegraphics[width=8cm]{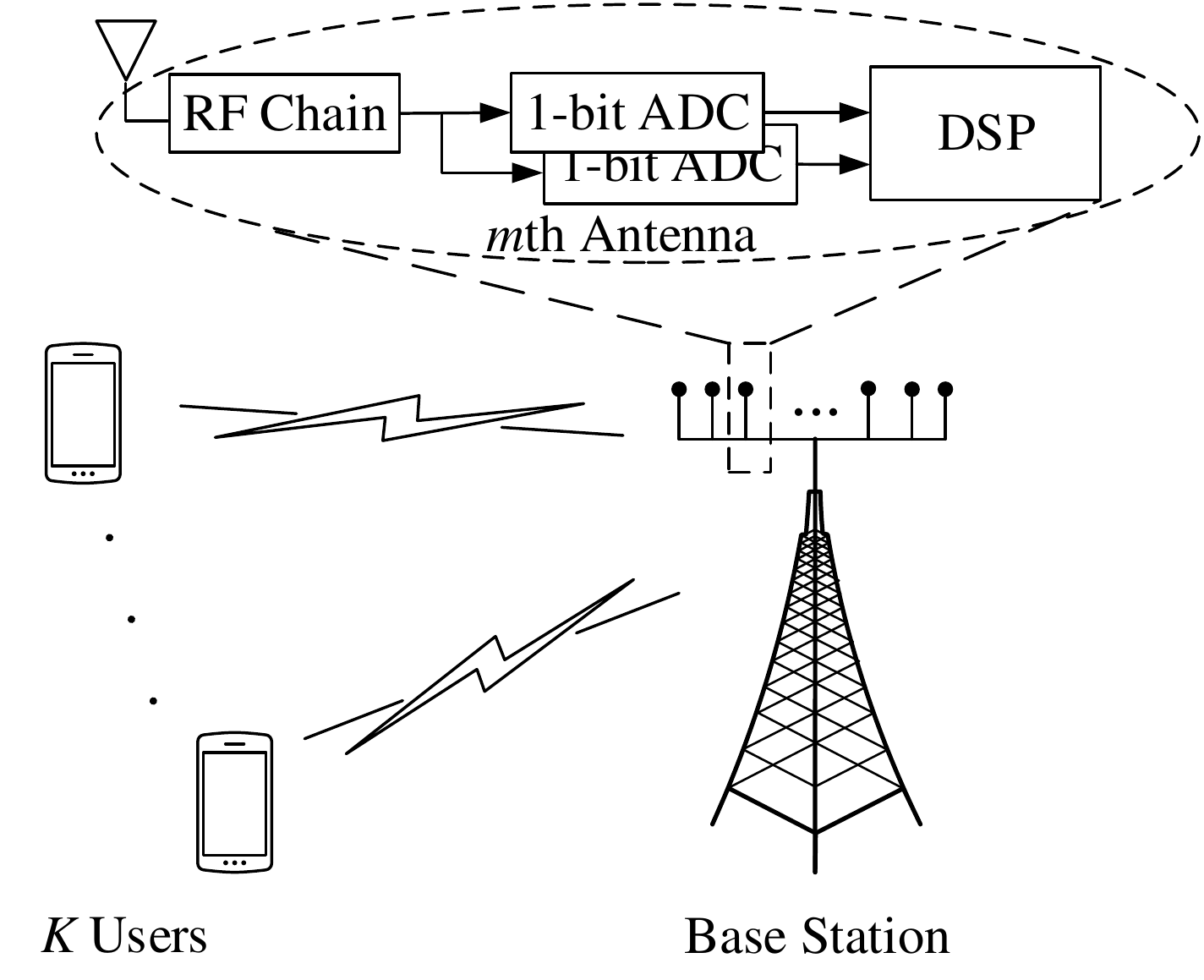}\\
  \vspace{-0.3cm}
  \caption{{\color{black}One-bit massive MIMO system architecture.}}\label{1bit}
  \vspace{-0.5cm}
\end{figure}

As depicted in Fig.~\ref{1bit}, we consider a single-cell one-bit massive MIMO system with $K$ single-antenna terminals and an $M$-antenna BS, where each antenna is equipped with two one-bit ADCs and $M\gg K\gg 1$ is assumed. For the uplink, we assume all $K$ users simultaneously transmit independent data symbols to the BS, so the received signal at the BS is
\begin{equation}
{\mathbf{y}} = \sqrt {{\rho_{\rm d}}} {\mathbf{Hs}} + {\mathbf{n}},
\end{equation}
where {\color{black}$\mathbf{n}\sim\mathcal{CN}({\mathbf{0}},{\color{black}\mathbf{I}_{M}})$} is the $M\times 1$ additive white Gaussian noise vector, {\color{black}$\mathbf{H}$} is the $M\times K$ channel matrix, and $\mathbf{s}$ is a vector containing the signals transmitted by each user. {\color{black} We also define the vectorized channel $\underline{\mathbf{h}} = {\rm vec}(\mathbf{H})$ and we assume that $\underline{\mathbf{h}}\sim\mathcal{CN}(\mathbf{0},\mathbf{C}_{\underline{\mathbf{h}}})$, where $\mathbf{C}_{\underline{\mathbf{h}}}$ is the covariance matrix of $\underline{\mathbf{h}}$}. We assume $\E\{|s_k|^2\} = 1$, and we will define the scale factor $\rho_{\rm d}$ to be the uplink SNR. {\color{black} Due to our assumption of one-bit quantization below and hence the lack of any signal dynamic range, we must assume that some type of power control is implemented that prevents a strong user from overwhelming other weaker users. For this reason, in our model we assume all users have the same level of large-scale fading/SNR $\rho_{\rm d}$.}

The quantized signal obtained after the one-bit ADCs is represented as
\begin{equation}\label{system_quantization}
\mathbf{r} = \mathcal{Q}(\mathbf{y}) = \mathcal{Q}\left(\sqrt {{\rho_{\rm d}}} {\mathbf{Hs}} + {\mathbf{n}}\right),
\end{equation}
where $\mathcal{Q}(.)$ represents the one-bit quantization operation, which is applied separately to the real and imaginary part as $\mathcal{Q}(.) = \frac{1}{\sqrt{2}} \left(\textrm{sign}\left(\Re\left(.\right)\right)+ j\textrm{sign}\left(\Im\left(.\right)\right)\right)$. Thus, the output set of the one-bit quantization is equivalent to the QPSK constellation points $\mathcal{R} = \frac{1}{\sqrt{2}}\{1+j, 1-j, -1+j, -1-j\}$.


\section{Channel Estimation for One-bit MIMO}
In a standard implementation, the CSI is estimated at the BS and then used to detect the data symbols transmitted from the $K$ users. In the uplink transmission phase, we assume the coherence interval is divided into two parts: one dedicated to training and the other to data transmission. During training, all $K$ users simultaneously transmit their pilot sequences of $\tau$ symbols each to the BS, which yields
\begin{equation}\label{received_signal_training}
\mathbf{Y}_{\rm p} = \sqrt{\rho_{\rm p}}\mathbf{H}\bm{\Phi}^T +\mathbf{N}_{\rm p},
\end{equation}
where $\mathbf{Y}_{\rm p}\in\mathbb{C}^{M\times\tau}$ is the received signal, $\rho_{\rm p}$ is the pilot transmit power, and $\bm{\Phi}\in\mathbb{C}^{\tau\times K}$ is the pilot matrix transmitted from the $K$ users. We assume all pilot sequences are column-wise orthogonal, i.e., $\bm{\Phi}^T\bm{\Phi}^* = \tau {\color{black}\mathbf{I}_{K}}$, which implies $\tau \ge K$. {\color{black} We further assume that both SNRs $\rho_{\rm d}$ and $\rho_{\rm p}$ are known at the BS via, for example, a low-rate control channel.}

To match the matrix form of \eqref{received_signal_training} to the vector form of \eqref{system_quantization}, we vectorize the received signal as
\begin{equation}\label{vec_received_signal_training}
\vect(\mathbf{Y}_{\rm p}) = \mathbf{y}_{\rm p} = \bar{\bm{\Phi}}\underline{\mathbf{h}} + \underline{\mathbf{n}}_{\rm p},
\end{equation}
where $\bar{\bm{\Phi}} = \left(\bm{\Phi} \otimes \sqrt{\rho_{\rm p}}{\color{black}\mathbf{I}_{M}}\right)$ and $\underline{\mathbf{n}}_p = \vect(\mathbf{N}_p)$. After one-bit ADCs, the quantized signal can be expressed as
\begin{equation}\label{Training_Quantized_Signal}
  \mathbf{r}_{\rm p} = \mathcal{Q}(\mathbf{y}_{\rm p}),
\end{equation}
where the $i$th element of $\mathbf{r}_{\rm p}$ takes values from the set $\mathcal{R}$.

\subsection{Bussgang-Based Channel Estimator}

The authors in \cite{chiara2014massive,juncil2015near,jacobsson2015one,jianhua2014channel} have investigated various methods for channel estimation in one-bit systems that rely on either the maximum-likelihood algorithm or on iterative algorithms with relatively high complexity. Furthermore, the channel estimators obtained by these methods do not lend themselves to an analysis that provides insight on their performance.

To address these drawbacks, in this section we take a more fundamental approach and derive simple linear estimators whose performance can be analyzed in a straightforward way.  These estimators are based on the so-called Bussgang decomposition \cite{bussgang1952yq}, which finds a statistically equivalent (up to first and second moments) linear operator for any nonlinear function of a Gaussian signal.  In particular, for the one-bit quantizer in~(\ref{Training_Quantized_Signal}), the Bussgang decomposition is written
\begin{equation}\label{eq:Bussgang}
  \mathbf{r}_{\rm p} = \mathcal{Q}(\mathbf{y}_{\rm p}) = \mathbf{A}_{\rm p} \mathbf{y}_{\rm p} + \mathbf{q}_{\rm p} \; ,
\end{equation}
where $\mathbf{A}_{\rm p}$ is the linear operator and $\mathbf{q}_{\rm p}$ the statistically equivalent quantizer noise. The matrix $\mathbf{A}_{\rm p}$ is chosen to make $\mathbf{q}_{\rm p}$ uncorrelated with $\mathbf{y}_{\rm p}$ \cite{mezghani2012capacity,bussgang1952yq}, or equivalently, to minimize the power of the equivalent quantizer noise.  This yields
\begin{equation}\label{optimal_A_p}
\mathbf{A}_{\rm p} = \mathbf{C}_{\mathbf{y}_{\rm p}\mathbf{r}_{\rm p}}^{H}\mathbf{C}_{\mathbf{y}_{\rm p}}^{-1},
\end{equation}
where $\mathbf{C}_{\mathbf{y}_{\rm p}\mathbf{r}_{\rm p}}$ denotes the cross-correlation matrix between the received signal $\mathbf{y}_{\rm p}$ and the quantized signal $\mathbf{r}_{\rm p}$, and $\mathbf{C}_{\mathbf{y}_{\rm p}}$ denotes the auto-correlation matrix of $\mathbf{y}_{\rm p}$. For one-bit quantization {\color{black}and Gaussian inputs}, $\mathbf{C}_{\mathbf{y}_{\rm p}\mathbf{r}_{\rm p}}$ is given by \cite{bussgang1952yq}\cite[Ch.10]{papoulis2002probability}
\begin{align}\label{Cross_Quantize_Receive}
{{\bf{C}}_{{\mathbf{y}_{\rm p}\mathbf{r}_{\rm p}}}} & = \sqrt {\frac{2}{\pi }} {{\bf{C}}_{{\mathbf{y}_{\rm p}}}}{\rm{diag}}{\left( {{{\bf{C}}_{{\mathbf{y}_{\rm p}}}}} \right)^{ - \frac{1}{2}}} \triangleq \sqrt {\frac{2}{\pi }} {{\bf{C}}_{{\mathbf{y}_{\rm p}}}}\bm{\Sigma}_{\mathbf{y}_{\rm p}}^{-\frac{1}{2}}.
\end{align}
where $\bm{\Sigma}_{\mathbf{y}_{\rm p}} = {\rm{diag}}\left( {{{\bf{C}}_{{\mathbf{y}_{\rm p}}}}} \right)$.

Using~\eqref{vec_received_signal_training} and~\eqref{eq:Bussgang}, we can express $\mathbf{r}_{\rm p}$ as
\begin{equation}\label{Training_Quantized_Signal_MMSE}
  \mathbf{r}_{\rm p} = \mathcal{Q}(\mathbf{y}_{\rm p}) = \tilde{\bm{\Phi}}\underline{\mathbf{h}} + \tilde{\mathbf{n}}_{\rm p},
\end{equation}
where $\tilde{\bm{\Phi}} = \mathbf{A}_{\rm p}{\bar{\bm{\Phi}}} \in\mathbb{C}^{M\tau\times M\tau}$,  $\tilde{\mathbf{n}}_{\rm p} =\mathbf{A}_{\rm p}\mathbf{n}_{\rm p} + \mathbf{q}_{\rm p}\in\mathbb{C}^{M\tau\times 1}$. {\color{black}For the sake of simplicity, we derive the subsequent formulas for the case of $\mathbf{C}_{\underline{\mathbf{h}}}=\mathbf{I}_{MK}$. We will show later in Section VI.A that they are readily modified to include a generic $\mathbf{C}_{\underline{\mathbf{h}}}$.}

The matrix
$\mathbf{A}_{\rm p}$ is given by substituting~\eqref{Cross_Quantize_Receive} into~\eqref{optimal_A_p}:{\color{black}
\begin{align}\label{A_p1}
{{\bf{A}}_p} & = \sqrt {\frac{2}{\pi }} \diag\left(\mathbf{C}_{\mathbf{y}_{\rm p}}\right)^{-\frac{1}{2}} \nonumber \\
&= \sqrt {\frac{2}{\pi }} \diag\left(\left(\bm{\Phi}\bm{\Phi}^H \otimes \rho_{\rm p}{\color{black}\mathbf{I}_{M}}\right) + {\color{black}\mathbf{I}_{M\tau}}\right)^{-\frac{1}{2}}.
\end{align}}

{\color{black} We can see from \eqref{A_p1} that $\mathbf{A}_{\rm p}$ depends on the specific choice of pilot sequences ${\bm{\Phi}}$. In order to obtain a simple expression for $\mathbf{A}_{\rm p}$, we will consider pilot sequences composed of submatrices of the discrete Fourier transform (DFT) operator \cite{biguesh2004downlink}. {\color{black} In particular, we define $\bm{\Phi}$ using $K$ columns of the $\tau\times \tau$ DFT matrix, in which case $\bm{\Phi}$ has dimension $\tau\times K$, where $\tau \ge K$.} The benefits of using DFT pilot sequences are: i) all the elements of the matrix have the same magnitude, which simplifies peak transmit power constraints, and ii) the diagonal terms of $\bm{\Phi}\bm{\Phi}^H$ are always equal to $K$, which results in a simple expression for $\mathbf{A}_{\rm p}$, as follows:}
\begin{equation}\label{A_t2}
\mathbf{A}_{\rm p} = \sqrt{\frac{2}{\pi}\frac{1}{K\rho_{\rm p} + 1}}{\color{black}\mathbf{I}_{M\tau}} \triangleq \alpha_{\rm p}{\color{black}\mathbf{I}_{M\tau}}.
\end{equation}

Based on this statistically equivalent linear model, we can formulate the LMMSE estimator \cite{kay1993fundamentals}, which we refer to as Bussgang LMMSE (BLMMSE) channel estimator:
\begin{align}
\label{BMMSE_Channel}\hat{\underline{\mathbf{h}}}^{\texttt{BLM}} = \mathbf{C}_{\underline{\mathbf{h}}\mathbf{r}_{\rm p}} \mathbf{C}_{\mathbf{r}_{\rm p}}^{-1}\mathbf{r}_{\rm p}&= \left(\tilde{\bm{\Phi}}^H+\mathbf{C}_{\underline{\mathbf{h}}\mathbf{q}_{\rm p}}\right) \mathbf{C}_{\mathbf{r}_{\rm p}}^{-1}\mathbf{r}_{\rm p},
\end{align}
where $\mathbf{C}_{\mathbf{h}\mathbf{r}_{\rm p}}$ is the cross-correlation matrix between $\mathbf{h}$ and $\mathbf{r}_{\rm p}$, $\mathbf{C}_{\mathbf{r}_{\rm p}}$ is the auto-correlation matrix of $\mathbf{r}_{\rm p}$.

The formula of \eqref{BMMSE_Channel} involves the auto-correlation function of the quantized signal $\mathbf{r}_{\rm p}$. It has been shown in \cite{jacovitti1994estimation} that for one-bit ADCs, the arcsin law can be used to obtain
\begin{align}\label{C_rr}
\mathbf{C}_{\mathbf{r}_{\rm p}} =& \frac{2}{\pi} \left(\arcsin\left(\bm{\Sigma}^{-\frac{1}{2}}_{\mathbf{y}_{\rm p} } \Re\left({\mathbf{C}_{\mathbf{y}_{\rm p} }}\right) \bm{\Sigma}^{-\frac{1}{2}}_{\mathbf{y}_{\rm p} } \right) \right. \nonumber \\
& \left. + j \arcsin\left(\bm{\Sigma}^{-\frac{1}{2}}_{\mathbf{y}_{\rm p} } \Im\left({\mathbf{C}_{\mathbf{y}_{\rm p} }}\right) \bm{\Sigma}^{-\frac{1}{2}}_{\mathbf{y}_{\rm p} } \right)  \right).
\end{align}
Moreover, using $\mathbf{A}_{\rm p}$ as in~\eqref{A_p1} according to the Bussgang theorem, the quantizer noise $\mathbf{q}_{\rm p}$ is not only uncorrelated with the received signal $\mathbf{y}_{\rm p}$, but also with the channel $\underline{\mathbf{h}}$ (see Appendix \ref{proof_uncorrelation}). Therefore, we can simplify the BLMMSE channel estimator of \eqref{BMMSE_Channel} as
\begin{align}\label{BMMSE_Channel_Approx_Orig}
\underline{\hat{\mathbf{h}}}^{\texttt{BLM}} = \tilde{\bm{\Phi}}^H\mathbf{C}_{\mathbf{r}_{\rm p}}^{-1}\mathbf{r}_{\rm p}.
\end{align}

Thus the covariance matrix of the BLMMSE channel estimate is given by
\begin{equation}\label{MSE_BLMMSE_Exact}
\mathbf{C}_{\underline{\hat{\mathbf{h}}}^{\texttt{BLM}}} = \tilde{\bm{\Phi}}^H\mathbf{C}_{\mathbf{r}_{\rm p}}^{-1}\tilde{\bm{\Phi}}.
\end{equation}

{\color{black} A similar LMMSE channel estimator is proposed in \cite{mollen2016performance}. However, our proposed channel estimator in \eqref{BMMSE_Channel_Approx_Orig} is more general since the correlation between each element of the quantizer noise is taken into account by using the arcsine law. In fact, \eqref{MSE_BLMMSE_Exact} can be reduced to the estimator derived in \cite{mollen2016performance} if $\tau = K$ is assumed. When $\tau = K$, it is easy to see that $\mathbf{C}_{\mathbf{y}_{\rm p}}=(K\rho_{\rm p}+1){\color{black}\mathbf{I}_{MK}}$, and hence according to \eqref{C_rr}, $\mathbf{C}_{\mathbf{r}_{\rm p}}={\color{black}\mathbf{I}_{MK}}$. Therefore, when $\tau=K$ we can obtain the BLMMSE channel estimator of \eqref{BMMSE_Channel_Approx_Orig} as simply
\begin{align}\label{BMMSE_Channel_Approx}
\underline{\hat{\mathbf{h}}}^{\texttt{BLM}} = \tilde{\bm{\Phi}}^H\mathbf{r}_{\rm p}.
\end{align}

We emphasize that, when $\tau=K$, there is no correlation between the quantizer noise $\mathbf{q}_{\rm p}$ and the normalized MSE for BLMMSE channel estimator is given by
\begin{align}\label{MSE_MMSE}
{{\cal M}}^{\texttt{BLM}} &= \frac{1}{MK}\E\left\{\left\|\tilde{\bm{\Phi}}^H\mathbf{r}_{\rm p} - \underline{\mathbf{h}} \right\|_2^2\right\} \nonumber \\
&= \frac{1}{MK}\tr\left( {\color{black}\mathbf{I}_{MK}} - \tilde{\bm{\Phi}}^H \tilde{\bm{\Phi}} \right) \nonumber \\
&= 1 - \frac{2K\rho_{\rm p}}{\pi( K\rho_{\rm p} +1)} \; ,
\end{align}
and for high SNRs
\begin{equation}\label{Limit_MSE_MMSE}
\mathop {\lim }\limits_{{\rho _p} \to \infty } {{\cal M}}^{\texttt{BLM}} = 1 - \frac{2}{\pi} = -4.40 \textrm{dB}.
\end{equation}
The results in \eqref{MSE_MMSE} and \eqref{Limit_MSE_MMSE} are allied with the results in \cite[Eq.(35)]{mollen2016performance} by setting $p[l] = 1/L$ and $\beta_k P_k = \rho_{\rm p}$. In addition, the result in \eqref{Limit_MSE_MMSE} implies that there exists an error floor for the channel estimate as the training power increases to infinity.}

{
\subsection{Extension to Frequency Selective Fading with OFDM}
Although for simplicity we focus on the flat fading case in this paper, we show here how to extend our channel estimation method to the frequency selective case, assuming the transmitter employs OFDM signaling. In particular, consider an OFDM system with $N_{\rm c}$ subcarriers, and denote the uplink OFDM symbol transmitted from the $k$th user as $\mathbf{x}_k^{\rm FD}\in\mathbb{C}^{N_{\rm c}\times 1}$. Before transmission, this vector is processed by a unitary IFFT operation $\mathbf{F}^H$, and then a cyclic prefix (CP) of length $N_{\rm cp}$ is added. Assume the CP length satisfies $L-1\le N_{\rm cp}\le N_{\rm c}$, where $L$ is the number of channel taps. After removing the CP, the $N_{\rm c}\times 1$ received time domain signal at the $m$th BS-antenna is given by
\begin{align}\label{received_signal_mth}
\mathbf{y}_{m}^{\rm TD} &= \sum_{k=1}^K\mathbf{G}_{mk}^{\rm TD}\mathbf{F}^H\mathbf{x}_k^{\rm FD} + \mathbf{n}_m^{\rm TD} \nonumber \\
& = \sum_{k=1}^K \bm{\Phi}_k^{\rm TD} \mathbf{g}_{mk}^{\rm TD} + \mathbf{n}_{m}^{\rm TD} = \sum_{k=1}^K \bm{\Phi}_{k,L}^{\rm TD} \mathbf{h}_{mk}^{\rm TD} + \mathbf{n}_{m}^{\rm TD},
\end{align}
where the superscripts ``TD'' and ``FD'' refer to Time Domain and Frequency Domain, respectively. The matrix $\mathbf{G}_{mk}^{\rm TD}\in\mathbb{C}^{N_{\rm c}\times N_{\rm c}}$ is circulant and its first column is given by $\mathbf{g}_{mk}^{\rm TD}=[(\mathbf{h}_{mk}^{\rm TD})^T,0,...,0]^T$, where $\mathbf{h}_{mk}^{\rm TD}$ is an $L\times 1$ column vector containing the $L$ channel taps, and $\mathbf{n}_m^{\rm TD}\sim\mathcal{CN}(\mathbf{0},\mathbf{I})$ is additive white Gaussian noise. The matrix $\bm{\Phi}_k^{\rm TD}\in\mathbb{C}^{N_{\rm c}\times N_{\rm c}}$ is also circulant with first column given by $\bm{\phi}_k^{\rm TD}= \mathbf{F}^H\mathbf{x}_k^{\rm FD}$. $\bm{\Phi}_{k,L}^{\rm TD}$ is a submatrix of $\bm{\Phi}_{k}^{\rm TD}$, corresponding to the first $L$ columns of $\bm{\Phi}_{k}^{\rm TD}$.  The second equation follows from the commutative property of circulant convolution. The third equation is due to the fact that there are only finite $L$ channel taps.

After stacking the received time domain signal $\mathbf{y}_{m}^{\rm TD}$ for all $M$ BS antennas, we have
\begin{equation}
\mathbf{y}^{\rm TD} = \bar{\bm{\Phi}}_{L}^{\rm TD}\mathbf{h}^{\rm TD} + \mathbf{n}^{\rm TD},
\end{equation}
where $\bar{\bm{\Phi}}_{L}^{\rm TD} = \mathbf{I}\otimes \bm{\Phi}_{L}^{\rm TD}$ with $\bm{\Phi}^{\rm TD}_{L} = [\bm{\Phi}_{1,L}^{\rm TD},\bm{\Phi}_{2,L}^{\rm TD},...,\bm{\Phi}_{K,L}^{\rm TD}]\in\mathbb{C}^{N_{\rm c}\times LK}$ and $\mathbf{h}^{\rm TD}\in\mathbb{C}^{MKL\times 1}$ contains all channel taps between the $M$ BS-antennas and $K$ users. After one-bit quantization, the time domain quantized signal can be expressed as
\begin{equation}\label{quantized_signal_CE}
\mathbf{r}^{\rm TD} = \mathcal{Q}(\mathbf{y}^{\rm TD}) = \mathcal{Q}(\bar{\bm{\Phi}}_{L}^{\rm TD}\mathbf{h}^{\rm TD} + \mathbf{n}^{\rm TD}) \; ,
\end{equation}
and we see that, unlike a conventional system, OFDM cannot split the wideband channel into many parallel narrowband channel in a one-bit system.

Using the Bussgang decomposition, the non-linear quantization operation can be reformulated as
\begin{align}
\mathbf{r}^{\rm TD} &= \mathbf{A}\mathbf{y}^{\rm TD} + \mathbf{q}^{\rm TD} \nonumber \\
& = \mathbf{A} \bar{\bm{\Phi}}_{L}^{\rm TD}\mathbf{h}^{\rm TD}  + \mathbf{A}\mathbf{n}^{\rm TD}+ \mathbf{q}^{\rm TD},
\end{align}
where the matrix  $\mathbf{A}$ is chosen to make the quantizer noise $\mathbf{q}^{\rm TD}$ uncorrelated with $\mathbf{y}^{\rm TD}$. If the received time domain signal $\mathbf{y}^{\rm TD}$ is Gaussian, we have
\begin{align}
\mathbf{A} &= \sqrt{\frac{2}{\pi}}\diag\left(\mathbf{C}_{\mathbf{y}^{\rm TD}}\right)^{-\frac{1}{2}} \; .
\end{align}
Consequently, the BLMMSE channel estimator for the wideband OFDM case can be expressed as
\begin{align}
\hat{\mathbf{h}}^{\rm TD} = \mathbf{C}_{\mathbf{h}^{\rm TD}}(\bar{\bm{\Phi}}_{L}^{\rm TD})^H\mathbf{A}^H\mathbf{C}_{\mathbf{r^{\rm TD}}}^{-1}\mathbf{r}^{\rm TD},
\end{align}
where $ \mathbf{C}_{\mathbf{h}^{\rm TD}}$ is the covariance matrix of ${\mathbf{h}^{\rm TD}}$, and the covariance matrix of $\mathbf{r}^{\rm TD}$ is obtained by using the arcsine law:
\begin{align}
\mathbf{C}_{\mathbf{r}^{\rm TD}} =& \frac{2}{\pi} \left(\arcsin\left(\bm{\Sigma}^{-\frac{1}{2}}_{\mathbf{y}^{\rm TD}} \Re\left({\mathbf{C}_{\mathbf{y}^{\rm TD}}}\right) \bm{\Sigma}^{-\frac{1}{2}}_{\mathbf{y}^{\rm TD}} \right)\right.  \nonumber \\
&\left.+ j \arcsin\left(\bm{\Sigma}^{-\frac{1}{2}}_{\mathbf{y}^{\rm TD}} \Im\left({\mathbf{C}_{\mathbf{y}^{\rm TD}}}\right) \bm{\Sigma}^{-\frac{1}{2}}_{\mathbf{y}^{\rm TD}} \right)  \right),
\end{align}
where $\bm{\Sigma}_{\mathbf{y}^{\rm TD}} = {\rm{diag}}\left( {{{\bf{C}}_{{\mathbf{y}^{\rm TD}}}}} \right)$. The covariance matrix of the quantizer noise $\mathbf{q}^{\rm TD}$ can be obtained by
\begin{equation}
\mathbf{C}_{\mathbf{q}^{\rm TD}} = \mathbf{C}_{\mathbf{r}^{\rm TD}} - \mathbf{A}\mathbf{C}_{\mathbf{y}^{\rm TD}}\mathbf{A}^H.
\end{equation}

The above Bussgang-based channel estimators are more general than those derived in other work such as \cite{mollen2016performance}, since they take into account the fact that in general the covariance matrix of the quantizer noise $\mathbf{q}^{\rm TD}$ cannot be expressed as a diagonal matrix due to the arcsine law. This observation holds for any linear modulation scheme employed by the users, not just OFDM. While the derivations that follow will focus on the flat fading case, we can see from the above that they can be easily generalized to frequency-selective fading.
}

{\color{black}
\subsection{Low SNR Approximate BLMMSE Channel Estimate Covariance}

As we can see from \eqref{C_rr} and \eqref{BMMSE_Channel_Approx_Orig}, it is difficult to obtain a general closed-form expression for the MSE of the BLMMSE channel estimator due to the `arcsine' operation. However, it is expected that massive MIMO systems will operate at relatively low SNRs due to the availability of a large array gain \cite{ngo2013energy}. Therefore in this subsection, we focus on deriving a low-SNR approximation for the covariance matrix of the BLMMSE channel estimator. According to \eqref{Training_Quantized_Signal_MMSE}, we can reformulate $\mathbf{C}_{\mathbf{r}_{\rm p}}$ as the following linear function,
\begin{equation}\label{Linear_C_rr}
\mathbf{C}_{\mathbf{r}_{\rm p}} = \tilde{\bm{\Phi}}\tilde{\bm{\Phi}}^H + \mathbf{A}_{\rm p}\mathbf{A}_{\rm p}^H +\mathbf{C}_{\mathbf{q}_{\rm p}},
\end{equation}
where
\begin{align}\label{C_qq}
\mathbf{C}_{\mathbf{q}_{\rm p}} &= \mathbf{C}_{\mathbf{r}_{\rm p}} - \mathbf{A}_{\rm p}\mathbf{C}_{\mathbf{y}_{\rm p}}\mathbf{A}_{\rm p}^H \nonumber \\
& = \frac{2}{\pi}(\arcsin(\mathbf{X})+j\arcsin(\mathbf{Y})) - \frac{2}{\pi}(\mathbf{X} + j\mathbf{Y}),
\end{align}
and where we define
\begin{align}
\mathbf{X} &= \bm{\Sigma}^{-\frac{1}{2}}_{\mathbf{y}_{\rm p} } \Re\left({\mathbf{C}_{\mathbf{y}_{\rm p} }}\right)\bm{\Sigma}^{-\frac{1}{2}}_{\mathbf{y}_{\rm p}} \\
\mathbf{Y} &= \bm{\Sigma}^{-\frac{1}{2}}_{\mathbf{y}_{\rm p} } \Im\left({\mathbf{C}_{\mathbf{y}_{\rm p}}}\right)\bm{\Sigma}^{-\frac{1}{2}}_{\mathbf{y}_{\rm p}}.
\end{align}

We can see from \eqref{C_qq} that the covariance matrix of the quantizer noise is in general not a diagonal matrix, which implies that there exists correlation between the quantization noise on each antenna. However, at low SNR or for large numbers of users, $\mathbf{C}_{\mathbf{y}_{\rm p}}$ is diagonally dominant and we can use the following approximation for applying the arcsine law:
\begin{equation}
\frac{2}{\pi}\arcsin(a) \cong \left\{ \begin{array}{*{20}{c}}
{1,}&{a = 1}\\
{2a/\pi,}&{a<1}.
\end{array}
\right.
\end{equation}
Since the non-diagonal elements of $\mathbf{X}$ and $\mathbf{Y}$ are much smaller than 1 in the low SNR regime, we can approximate \eqref{C_qq} as
\begin{equation}\label{C_qq_app}
\mathbf{C}_{\mathbf{q}_{\rm p}} \cong (1-2/\pi){\color{black}\mathbf{I}_{M\tau}}.
\end{equation}
This implies that we can approximate the quantizer noise as uncorrelated noise with a variance of $1-2/\pi$ at low SNR. Substituting \eqref{C_qq_app} and \eqref{Linear_C_rr} into \eqref{MSE_BLMMSE_Exact}, we have
\begin{align}\label{BLM_MSE_App}
&\mathbf{C}_{\underline{\hat{\mathbf{h}}}^{\texttt{BLM}}}\cong \tilde{\bm{\Phi}}^H\left(\tilde{\bm{\Phi}}\tilde{\bm{\Phi}}^H+(\alpha_{\rm p}^2+1-2/\pi){\color{black}\mathbf{I}_{M\tau}}\right)^{-1}\tilde{\bm{\Phi}} \nonumber \\
& = (\alpha_{\rm p}^2 \tau\rho_{\rm p} + \alpha_{\rm p}^2+1-2/\pi)^{-1}\alpha_{\rm p}^2 \tau\rho_{\rm p}{\color{black}\mathbf{I}_{MK}} {\color{black}\triangleq } \sigma^2{\color{black}\mathbf{I}_{MK}} \; ,
\end{align}
{\color{black} where we have defined $\sigma^2 = (\alpha_{\rm p}^2 \tau\rho_{\rm p} + \alpha_{\rm p}^2+1-2/\pi)^{-1}\alpha_{\rm p}^2 \tau\rho_{\rm p}$.}
The equation on the second line holds due to the matrix inversion identity $(\mathbf{I} + \mathbf{AB})^{-1}\mathbf{A} = \mathbf{A}(\mathbf{I} + \mathbf{BA})^{-1}$. The result in \eqref{BLM_MSE_App} implies that in the low SNR regime, each element of the BLMMSE channel estimate is uncorrelated. In what follows, we will evaluate the uplink achievable rate by using the low SNR approximation in \eqref{BLM_MSE_App}. }


\section{Achievable Rate Analysis \\in the One-Bit MIMO Uplink}
\subsection{Data Transmission}
In the data transmission stage, we assume the $K$ users simultaneously transmit their data symbols, represented as the vector $\mathbf{s}$, to the BS. After one-bit quantization, the signal at the BS can be expressed as
\begin{align}\label{r_expression}
\mathbf{r} _d &= \mathcal{Q}(\mathbf{y}_{\rm d})=\mathcal{Q}(\sqrt{\rho_{\rm d}}\mathbf{H}\mathbf{s} +  \mathbf{n}_d)  \nonumber \\
&= \sqrt{\rho_{\rm d}}\mathbf{A}_{\rm d} \mathbf{H}\mathbf{s} + \mathbf{A}_{\rm d}\mathbf{n}_{\rm d} + \mathbf{q}_{\rm d},
\end{align}
where the same definitions as in the previous sections apply, but replacing the subscript `p' with `d', since the power $\rho_{\rm d}$ during data transmission may be different than during training.  Again, according to the Bussgang decomposition and {\color{black}assuming a Gaussian input}, we have
\begin{align}\label{A_d_General}
\mathbf{A}_{\rm d} =\sqrt {\frac{2}{\pi }} \diag\left(\mathbf{C}_{\mathbf{y}_{\rm d}}\right)^{-\frac{1}{2}} = \sqrt {\frac{2}{\pi }} \diag\left(\rho_{\rm d} \mathbf{H}\mathbf{H}^H + {\color{black}\mathbf{I}_{M}}\right)^{-\frac{1}{2}}.
\end{align}
{\color{black}
Note that, in contrast to the model of \cite{mollen2016performance}, in which the quantizer noise can still be correlated with the desired signal since the same Bussgang decomposition is employed for different channel realizations, the Bussgang decomposition in \eqref{A_d_General} is employed for each individual channel realization. This approach ensures that the quantizer noise is uncorrelated with the desired signal.

As can be seen in \eqref{A_d_General}, the covariance matrix $\mathbf{C}_{\mathbf{y}_{\rm d}}$ of the quantizer input (and hence, the matrix $\mathbf{H}\mathbf{H}^H$) must be known at the BS in order to implement the Bussgang decomposition. In practice, however, we can use the same technique provided in \cite{bar2002doa} to reconstruct the covariance matrix of $\mathbf{C}_{\mathbf{y}_{\rm d}}$ using the measurements of the quantizer output. In addition, relying on
channel hardening for $K \gg 1$ in massive MIMO systems and for {i.i.d.} unit-variance channel coefficients, we can approximate the matrix $\mathbf{A}_{\rm d}$ as
\begin{equation}\label{A_d}
\mathbf{A}_{\rm d} \cong \sqrt{\frac{2}{\pi}}\sqrt{\frac{1}{1+K\rho_{\rm d}}}{\color{black}\mathbf{I}_{M}} = \alpha_{\rm d} {\color{black}\mathbf{I}_{M}},
\end{equation}
without requiring perfect CSI. This approximate gain matrix is assumed { without derivation} in other previous work such as \cite{mollen2016performance}.

}


Next we assume the BS uses the BLMMSE channel estimate to compute a linear receiver to detect the data symbols transmitted from the $K$ users. The linear receiver attempts to separate the quantized signal into $K$ streams by multiplying the signal by the matrix $\mathbf{W}^T$ as follows:
\begin{align}\label{Linear_Receiver}
{\bf{\hat s}} &= \mathbf{W}^T\mathbf{r}_{\rm d}\nonumber\\
&= \sqrt{\rho_{\rm d}}\mathbf{W}^T{{\bf{A}}_{\rm d}}(\hat{\mathbf{H}}\mathbf{s} + \bm{\mathcal{E}}\mathbf{s}) + \mathbf{W}^T{{\bf{A}}_{\rm d}}{\bf{n}}_{\rm d} + \mathbf{W}^T{\bf{q}}_{\rm d} \; ,
\end{align}
where {\color{black} ${\hat{\mathbf{H}}}=\rm{unvec}(\underline{\hat{\mathbf{h}}}^{\texttt{BLM}}) $ is the estimated channel matrix (unvec is the inverse of the vec operator in Eq.~\eqref{vec_received_signal_training})} and $\bm{\mathcal{E}} = \mathbf{H}-\hat{\mathbf{H}}$ denotes the channel estimation error. The $k$th element of $\hat{\mathbf{s}}$ is then used to decode the signal transmitted from the $k$th user:
\begin{align}\label{hat_s_k}
  {\hat s_k} =& {\sqrt {{\rho_{\rm d}}} {\mathbf{{w}}}_k^T{{\bf{A}}_{\rm d}}{\hat{\bf{h}}_k}{s_k}}+  {\sqrt {{\rho_{\rm d}}} {\mathbf{{w}}}_k^T \sum \nolimits_{i \ne k}^K {{\bf{A}}_{\rm d}}{\hat{\bf{h}}_i}{s_i}} \nonumber\\
  &+{\sqrt {{\rho_{\rm d}}} {\mathbf{{w}}}_k^T \sum \nolimits_{i =1}^K {{\bf{A}}_{\rm d}}\bm{\varepsilon}_i{s_i}} + {{\mathbf{{w}}}_k^T{{\bf{A}}_{\rm d}}{\bf{n}}_{\rm d}} + {{\mathbf{{w}}}_k^T{\bf{q}}_{\rm d}},
\end{align}
where {\color{black}$\mathbf{w}_k$, $\hat{\mathbf{h}}_k$ and $\bm{\varepsilon}_k$ are the $k$th columns of $\mathbf{W}$, $\hat{\mathbf{H}}$ and $\bm{\mathcal{E}}$, respectively.}

The last four terms in \eqref{hat_s_k} respectively correspond to user interference, channel estimation error, AWGN noise and quantizer noise. In our analysis, we will consider the performance of the common MRC and ZF receivers, defined by
\begin{align}
\label{MRC_Matrix}&\mathbf{W}^T_{\texttt{MRC}} = {\hat{\mathbf{H}}}^H \\
\label{ZF_Matrix}&\mathbf{W}^T_{\texttt{ZF}} =\left({\hat{\mathbf{H}}}^H{\hat{\mathbf{H}}}\right)^{-1}{\hat{\mathbf{H}}}^H,
\end{align}
respectively.


\subsection{{\color{black}Uplink Achievable {\color{black}Rate Approximation} at Low SNR}}
Although prior work has obtained expressions for the mutual information or the achievable rate of one-bit systems using the joint probability distribution of
the transmitted and received symbols \cite{jianhua2014channel,chiara2014massive,jacobsson2016throughput}, this approach does not result in easily computable or insightful expressions. To overcome this drawback, in this section we provide a simple closed-form expression for an approximation of the achievable rate for both MRC and ZF processing in the low SNR region.
{\color{black}
Using the same reasoning as in Section III-C, the covariance matrix of $\mathbf{q}_{\rm d}$ can be expressed as
\begin{equation}\label{C_qd}
\mathbf{C}_{\mathbf{q}_{\rm d}} = \mathbf{C}_{\mathbf{r}_{\rm d}} - \mathbf{A}_{\rm d}\mathbf{C}_{\mathbf{y}_{\rm d}}\mathbf{A}_{\rm d}^H,
\end{equation}
where $\mathbf{C}_{\mathbf{r}_{\rm d}}$ is the covariance matrix of $\mathbf{r}_{\rm d}$ and can be obtained using the arcsine law in \eqref{C_rr}. Note that, again, the covariance matrix of \eqref{C_qd} is in general not a diagonal matrix, which implies that there exists some correlations among the elements of $\mathbf{q}_{\rm d}$. For the special case where $\mathbf{C}_{\mathbf{y}_{\rm d}}=\rho_{\rm d} \mathbf{H}\mathbf{H}^H + {\color{black}\mathbf{I}_{M}}$ is diagonally dominant due to low SNR or for large $K$ with {i.i.d.} channels, then similar to the pilot phase, the approximation (\ref{C_qq_app}) can be used. 
}

Furthermore, while the quantizer noise ${\bf q}_{\rm d}$ is non-Gaussian, we can obtain a lower bound on the achievable rate by making the worst-case assumption \cite{hassibi2003how,diggavi2001the} that in fact it is Gaussian with the same covariance matrix in \eqref{C_qd}. Using this approach and~\eqref{hat_s_k}, the ergodic achievable rate of the one-bit MIMO uplink is lower bounded by~\eqref{ergodic_achievable_rate} shown on the next page.
\begin{figure*}[!t]
\normalsize
\setcounter{MYtempeqncnt}{\value{equation}}
\setcounter{equation}{41}
\begin{equation}\label{ergodic_achievable_rate}
\tilde{R}_k = \E\left\{\log_2\left(1+\frac{\rho_{\rm d} \left|\mathbf{w}_k^T{{\bf{A}}_{\rm d}}{\hat{\bf{h}}_k}\right|^2}{\rho_{\rm d}\sum_{i \ne k}^K\left|\mathbf{w}_k^T{{\bf{A}}_{\rm d}}{\hat{\bf{h}}_i}\right|^2 + \rho_{\rm d}\sum_{i =1}^K\left|\mathbf{w}_k^T{{\bf{A}}_{\rm d}}{\bm{\varepsilon}_i}\right|^2+ \left\|\mathbf{w}_k^T{{\bf{A}}_{\rm d}}\right\|^2 + \mathbf{w}_k^T \mathbf{C}_{\mathbf{q}_{\rm d}}\mathbf{w}_k^* }\right)\right\}
\end{equation}
\setcounter{equation}{42}
\hrulefill
\vspace*{-0.3cm}
\end{figure*}
In order to obtain a closed-form expression for the achievable rate, we rewrite the detected signal in \eqref{hat_s_k} as a known mean gain (which only depends on the channel distribution instead of the instantaneous channel) times the desired symbol plus an uncorrelated effective noise, as follows:
\begin{equation}
{\hat s_k} = {\rm{E}}\left\{ {\sqrt {{\rho_{\rm d}}} \mathbf{w}_k^T{{\bf{A}}_{\rm d}}{{\bf{h}}_k}} \right\}{s_k} + {{{\tilde n}}_{{\rm d},k}},
\end{equation}
where $\tilde{{n}}_{{\rm d},k}$ is the effective noise given by
\begin{align}\label{effective_noise}
{{\tilde n}}_{{\rm d},k} =& \left ( {\sqrt {{\rho_{\rm d}}} \mathbf{w}_k^T{{\bf{A}}_{\rm d}}{{\bf{h}}_k} - {\rm{E}}\left\{ {\sqrt {{\rho_{\rm d}}} \mathbf{w}_k^T{{\bf{A}}_{\rm d}}{{\bf{h}}_k}} \right\}} \right){s_k} \nonumber\\
&+ \sqrt {{\rho_{\rm d}}} \mathbf{w}_k^T\mathop \sum \limits_{i \ne k}^K {{\bf{A}}_{\rm d}}{{\bf{h}}_i}{s_i} + \mathbf{w}_k^T{{\bf{A}}_{\rm d}}{\bf{n}}_{\rm d} + \mathbf{w}_k^T{\bf{q}}_{\rm d}.
\end{align}


\textit{Lemma 1:} In a massive MIMO system with one-bit quantization and $\mathbf{A}_{\rm d} = \alpha_{\rm d}{\color{black}\mathbf{I}_{M}}$, the uplink achievable rate for the $k$th user at low SNR can be approximated by
\begin{equation}\label{Achievable_Rate}
R_k = \log_2 \left( 1 + \frac{\rho_{\rm d}\alpha_{\rm d}^2\left| {\E\left\{ {\mathbf{w}_k^T {\mathbf{h}_k}} \right\}} \right|^2}{\rho_{\rm d}\alpha_{\rm d}^2\Var\left({\mathbf{w}_k^T {{\bf{h}}_k}}\right) + \textrm{UI}_k + \textrm{AQN}_k }\right),
\end{equation}
where
\begin{align}
\textrm{UI}_k &= \rho_{\rm d}\alpha_{\rm d}^2\sum_{i\neq k }^K \E\left\{\left|\mathbf{w}_k^T  \mathbf{h}_i\right|^2\right\}\\
\textrm{AQN}_k &= \left(\alpha_{\rm d}^2+1-\frac{2}{\pi}\right)\E\left\{\left\|\mathbf{w}_k^T \right\|^2\right\}.
\end{align}

\begin{IEEEproof}
See Appendix \ref{proof_Lemma1}.
\end{IEEEproof}


The result in \eqref{Achievable_Rate} is obtained by approximating the effective noise as Gaussian. In a massive MIMO system, the effective noise is a sum of a very large number of independent zero-mean terms, and thus we expect via the central limit theorem that the approximation will be asymptotically tight to the lower bound of \eqref{ergodic_achievable_rate} in $M$.
In Section V, it will be shown that the gap between the achievable rate approximation given by \eqref{Achievable_Rate} and the lower bound of the ergodic achievable rate given in \eqref{ergodic_achievable_rate} is small, which implies that our resulting closed-form expression is an excellent predictor of the system performance.

Based on Lemma 1, we derive in the theorems below closed-form expressions for the lower bound on the achievable rate for the MRC and ZF receivers.

{\textit{Theorem 1:}} For the MRC receiver with CSI estimated by the BLMMSE channel estimator, the achievable rate of the $k$th user in a one-bit massive MIMO uplink at low SNR can be approximated by
\begin{align}
R_{\texttt{MRC},k} &= {\log _2}\left( {1 + \frac{{{\rho_{\rm d}}\alpha_{\rm d}^2M \sigma^2}}{{{\rho_d}\alpha_{\rm d}^2 K + \alpha_{\rm d}^2 + \left( {1 - 2/\pi } \right)}}} \right) \nonumber \\
\label{MRC_Achievable_Rate}&={\log_2}\left( {1 + {{{\rho_{\rm d}}\alpha_{\rm d}^2M\sigma^2}}} \right).
\end{align}
\begin{IEEEproof}
See Appendix \ref{proof_theorem2}.
\end{IEEEproof}

{\color{black}
{\textit{Theorem 2:}} For the ZF receiver with CSI estimated by the BLMMSE channel estimator, the achievable rate of the $k$th user in a one-bit massive MIMO uplink at low SNR can be approximated by
\begin{equation}\label{ZF_Achievable_Rate}
R_{\texttt{ZF},k} = {\log _2}\left( {1 + \frac{{{\rho _{\rm d}}\alpha _{\rm d}^2\sigma^2(M-K)}}{{{\rho _{\rm d}}\alpha _{\rm d}^2 K \eta+ \alpha _{\rm d}^2 + \left( {1 - 2/\pi } \right)}}} \right),
\end{equation}
where $\eta = (1-\sigma^2)$.
}
\begin{IEEEproof}
See Appendix \ref{proof_theorem3}.
\end{IEEEproof}

\section{One-Bit Massive MIMO System Design}

The simple approximation for the achievable rate derived in the previous section provides us with a tool for easily quantifying the impact of system design decisions. In this section, we study design issues surrounding the length of the training sequence, the power allocated for training and data transmission, and the number of BS antennas. Our performance metric will be the sum spectral efficiency, defined by
\begin{equation}\label{Spectral_Efficiency}
\mathcal{S}_{\texttt{A}} = \frac{T-{\color{black}\tau}}{T} \sum_{k=1}^{K}{R_{\texttt{A},k}},
\end{equation}
where $T$ represents the length of the coherence interval, during which the channel satisfies the block fading model and stays constant. The notation $\texttt{A}\in\{\texttt{MRC},\texttt{ZF}\}$ indicates that we will perform the analysis for both the MRC and ZF receivers.

\subsection{Power Efficiency in One-Bit Massive MIMO}\label{sec:power}
In this section, we study the power efficiency achieved by one-bit massive MIMO systems, where an increase in the number of antennas can be traded for reduced transmit power at the user terminals. We will consider two cases: i) the training power $\rho_{\rm p}$ (and hence the channel estimation accuracy) is fixed, but user transmit power decreases as $1/M$; and ii) the training power $\rho_{\rm p}$ and data transmission power $\rho_{\rm d}$ are equal and scale as $1/\sqrt{M}$.

\subsubsection{Case I}
In the first case, we assume $\rho_{\rm p}$ is fixed and independent of $M$, while $\rho_{\rm d} = E_{\rm u}/M^c$ for a given $c$, where $E_{\rm u}$ is fixed independent of $M$. We will find the largest value for $c$ such that scaling down the users' power by $1/M^c$ results in no change in spectral efficiency as $M\to\infty$.
Substituting $\rho_{\rm d} = E_{\rm u}/M^c$ into \eqref{MRC_Achievable_Rate} and \eqref{ZF_Achievable_Rate} and assuming $M$ increases to infinity, we can readily see that choosing $c=1$ will result in the spectral efficiency converging to a fixed value. This implies that, when the channel estimation accuracy is fixed, the transmit power of each user can be reduced proportionally by $1/M$ for both the MRC and ZF receivers while maintaining a given sum spectral efficiency. Moreover, the asymptotic performance for MRC and ZF is the same and is given by
\begin{equation}\label{SE_Constant1}
{\color{black}\mathop{\lim }\limits_{M\rightarrow \infty} \mathcal{S}_{\texttt{A}}|_{{\rho _{\rm d}} = \frac{{{E_{\rm u}}}}{M}} = \frac{T-\tau}{T} K  \log_2\left(1+\frac{2}{\pi}{\sigma^2 E_{\rm u}}\right).}
\end{equation}

\subsubsection{Case II}
For the second case, we assume the training and data transmission power are reduced at the same rate: $\rho_{\rm p} = \rho_{\rm d} = E_{\rm u}/M^c$, where again $E_{\rm u}$ is fixed independent of $M$.
Substituting $\rho_{\rm p} = \rho_{\rm d} = E_{\rm u}/M^c$ into \eqref{MRC_Achievable_Rate} and \eqref{ZF_Achievable_Rate} and assuming $M$ increases to infinity, the value of $c=1/2$ can be seen to provide constant performance. Thus, we cannot reduce the user transmit power as aggressively as in the first case where the channel estimation accuracy is fixed. The asymptotic performance for MRC and ZF is again the same in this case, but with a different asymptotic value:
\begin{equation}\label{SE_Constant2}
{\color{black}\mathop{\lim }\limits_{M\rightarrow \infty} \mathcal{S}_{\texttt{A}}|_{{\rho _{\rm d}} = \rho_{\rm p} = \frac{{{E_{\rm u}}}}{\sqrt{M}}} =  \frac{T-\tau}{T} K  \log_2\left(1+\frac{4}{\pi^2}{\tau E_{\rm u}^2}\right).}
\end{equation}

Note that both of the spectral efficiency expressions in \eqref{SE_Constant1} and \eqref{SE_Constant2} are equivalent to that of $K$ SISO channels with transmit power {\color{black} ${2\sigma^2 E_{\rm u}}/\pi$ and ${4\tau E_{\rm u}^2}/\pi^2$}, respectively, without interference. Thus, even though one-bit ADCs are deployed at the BS, the spectral efficiency increases proportionally to the number of users $K$.

\subsection{{\color{black}Resource Allocation in One-Bit Massive MIMO System}}


It has been proved in \cite{hassibi2003how} that for conventional MIMO systems with infinite precision ADCs, the optimal training length is always $\tau = K$. {\color{black} However, due to the quantizer noise, we will see that this result does not hold for one-bit massive MIMO systems. Considerable gains in spectral efficiency can be obtained by proper resource allocation. Thus in this subsection, we assume the users can vary the training power and the data transmission power and study the  optimal resource allocation scheme that jointly selects the length of the training sequence, and the power allocated to training and data transmission with the goal of maximizing  the sum spectral efficiency.

Let $\rho$ be the average transmit power and $P= \rho T$ be the total power budget for the users in one coherence interval, which satisfies the constraint $\tau \rho_{\rm p} + (T-\tau)\rho_{\rm d} \le P$. Then, {\color{black} following the approach of \cite{ngo2014massive},} the optimization problem can be formulated as
\begin{align}
& \mathop{\text{maximize}}\limits_{\rho_{\rm p},\rho_{\rm d},\tau}
& & \mathcal{S}_{\texttt{A}} \nonumber\\
& \text{subject to} & &\tau \rho_{\rm p} + (T-\tau)\rho_{\rm d}\leq P \nonumber\\
\label{opt_BE} &&& K\le \tau \le T \\
&&&\rho_{\rm p}\ge 0, \rho_{\rm d}\ge 0.
\end{align}
For any power allocation in which the users do not employ the full energy budget, the users could increase their training power (and, thus, increase the channel estimation accuracy) without causing any inter-user interference in the data transmission phase, and hence in turn improve their rate. Therefore, we can replace the inequality constraint on the total energy budget with an equality constraint, i.e., $\tau\rho_{\rm p} + (T-\tau)\rho_{\rm d} = P$. To facilitate the presentation, let $\gamma\in(0,1)$ denote the fraction of the total energy budget that is devoted to pilot training, so that $\gamma P = \tau \rho_{\rm p}$ and $(1 - \gamma)P = (T - \tau)\rho_{\rm d}$.
The optimization problem in~\eqref{opt_BE} is then equivalent to
\begin{align}
& \mathop{\text{maximize}}\limits_{\gamma,\tau}
& & \mathcal{S}_{\texttt{A}}|_{\rho_{\rm p} = \frac{\gamma P}{\tau},\rho_{\rm d} = \frac{(1-\gamma) P}{T-\tau}} \nonumber\\ 
\label{opt_BE2} & \text{subject to} & & 0<\gamma<1, ~~K\le \tau \le T.
\end{align}

{\it{Lemma 2}}: For both the MRC and ZF receivers in one-bit massive MIMO, the optimal training length $\tau^*$ that maximizes the sum spectral efficiency is not always equal to the number of users.
\begin{IEEEproof}
See Appendix \ref{proof_Lemma2}.
\end{IEEEproof}

Although we cannot obtain a closed-form expression for $\tau^*$, we can numerically evaluate $\tau^*$ using a simple search algorithm since there are only a few parameters in problem \eqref{opt_BE2}. As we will show in the numerical results, unlike conventional MIMO systems, the optimal training duration depends on various system parameters such as the coherence interval $T$ and the total energy budget $P$.}

\subsection{How Many More Antennas are Needed for One-Bit Massive MIMO?}
In this subsection, we compare the performance of one-bit and conventional massive MIMO with infinite resolution ADCs in terms of the number of antennas deployed at the BS. In particular, we wish to answer the question of how many more antennas a one-bit massive MIMO system would need to achieve the same spectral efficiency of a conventional massive MIMO implementation. For this analysis, we denote the number of antennas in the one-bit and conventional massive MIMO systems as $M_{\rm one}$ and $M_{\rm conv}$, respectively, and show the lower bound on the uplink achievable rate for both one-bit and conventional MIMO in Table I, {\color{black}where we define $C(x)=\frac{T-\tau}{T} K\log_2(1+x)$}.

{\color{black} For the special case of $\tau = K$ and $\rho_{\rm d} = \rho_{\rm p}$, it was shown in \cite{mollen2016performance} that 2.5 times more antennas are needed in one-bit systems to ensure the same rate as the conventional system with MRC, and also for ZF at low SNR.  This can be easily verified using our results as well. However, this result will not hold in general for the optimal values of $\tau, \rho_{\rm d}$ and $\rho_{\rm p}$ resulting from the optimization in \eqref{opt_BE2}. In fact, we can pose a complementary optimization problem in which we attempt to minimize the ratio $\kappa = M_{\rm one}/M_{\rm conv}$ required for both systems to achieve the same spectral efficiency, as follows:
\begin{align}
& \mathop{\text{minimize}}\limits_{\gamma,\tau,\kappa} & & \kappa \nonumber\\ 
\label{opt_Ratio} & \text{subject to} & &  \mathcal{S}_{\texttt A}^{\rm one} =\mathcal{S}_{\texttt A}^{\rm conv}, \nonumber \\
& & & 0<\gamma<1, ~~~K\le \tau \le T.
\end{align}
where $\mathcal{S}_{\texttt A}^{\rm conv}$ is the maximum spectral efficiency achieved for the conventional system by optimizing $\rho_{\rm p}$ and $\rho_{\rm d}$ with $\tau =K$ for fixed $M_{\rm conv}$. Since the problem in \eqref{opt_Ratio} only has a few parameters, we can use a simple search algorithm for the optimization.

Although no closed-form expression for the optimal $\kappa$ can be obtained, we will show in the simulations that less than 2.5 times more antennas are needed for the MRC receiver, and also for the ZF receiver at low SNR, if the training length $\tau$, training power $\rho_{\rm p}$ and data transmission power $\rho_{\rm d}$ are all optimized.}

\begin{table}\renewcommand{\arraystretch}{1.5}
\caption{Lower bound on individual achievable rates for conventional and one-bit MIMO systems}\label{Table_C}
\begin{tabular}{|c|c|c|}
  \hline
    & Conv. MIMO \cite{ngo2013energy} & One-bit MIMO\\ \hline
  MRC & $C\left(\frac{\rho_{\rm d} \tau\rho_{\rm p} M_{\rm conv}}{(1+K\rho_{\rm d})(1+\tau\rho_{\rm p})}\right)$ & $C\left({{{\rho _{\rm d}}\alpha _{\rm d}^2M_{\rm one}\sigma^2}}\right) $\\ \hline
  ZF & $C\left(\frac{\rho_{\rm d} \tau\rho_{\rm p} (M_{\rm conv}-K)}{K\rho_{\rm d} +\tau \rho_{\rm p} + 1}\right)$ & $C\left(\frac{{{\rho _{\rm d}}\alpha _{\rm d}^2\sigma^2(M_{\rm one}-K)}}{{{\rho _{\rm d}}\alpha _{\rm d}^2 K \eta+ \alpha _{\rm d}^2 + \left( {1 - 2/\pi } \right)}}\right)$ \\ \hline
\end{tabular}
\vspace{-0.5cm}
\end{table}

\section{Numerical Results}
The simulation results presented here consider an uplink single-cell one-bit massive MIMO system with a coherence interval of $T = 200$ symbols. Unless otherwise indicated, we assume $\rho_{\rm p} = \rho_{\rm d} = \rm{SNR}$.

\subsection{Channel Estimation Performance}
In this subsection, we evaluate the performance of the BLMMSE channel estimator proposed in {Section III-A} compared with the LS channel estimator of \cite{chiara2014massive} and the near maximum-likelihood channel estimator of \cite{juncil2015near}. Note that although the nML channel estimator proposed in \cite{juncil2015near} focused on estimating the channel vector between the $K$ users and one receive antenna, we can define the nML estimator for the entire channel for all $M$ receive antennas and $K$ users using logic similar to \cite{juncil2015near} as follows:
\begin{align}
\hat{\mathbf{h}}^{\texttt{nML}} = \mathop {\arg \max }\limits_{\scriptstyle{\acute{\mathbf{h}}_{\rm R}}\in {\mathbb{R}^{2MK \times 1}}\hfill\atop
\scriptstyle{\left\| \acute{{\mathbf{h}}}_{\rm R} \right\|^2} \le K\hfill}  \sum_{i=1}^{2M\tau}{\log\left(F\left(\sqrt{2}\bar{\bm{\varphi}}_{\rm R}^{(i)} \acute{\mathbf{h}}_{\rm R} \right)\right)},
\end{align}
{where $F(x)$ is the cumulative distribution function (CDF) of the standard normal distribution, and $\bar{\bm{\varphi}}_{\rm R}^{(i)} = \sqrt{2}r_{\rm R,p}^{(i)} \bar{\mathbf{\Phi}}_{\rm R}^{(i)}$. $r_{\rm R,p}^{(i)}$ and $\bar{\mathbf{\Phi}}_{\rm R}^{(i)}$ are respectively the $i$th element of $\mathbf{r}_{\rm R,p}$ and the $i$th row of $\bar{\bm{\Phi}}_{\rm R}$:
\begin{equation}
\mathbf{r}_{\rm R,p} = \left[ \Re(\mathbf{r}_{\rm p}) ~~ \Im(\mathbf{r}_{\rm p})\right]^T
\end{equation}
\begin{equation}
\bar{\bm{\Phi}}_{\rm R} = \left[ {\begin{array}{cc}
{\Re\left( {\bar{\bm{\Phi}}} \right)}&{ - \Im\left( {\bar{\bm{\Phi}}} \right)}\\
\Im\left( {\bar{\bm{\Phi}}} \right)&{\Re\left( {\bar{\bm{\Phi}}} \right)}
\end{array}} \right] \; .
\end{equation}
}

\begin{figure}[!t]
  \centering
  \includegraphics[width=8cm]{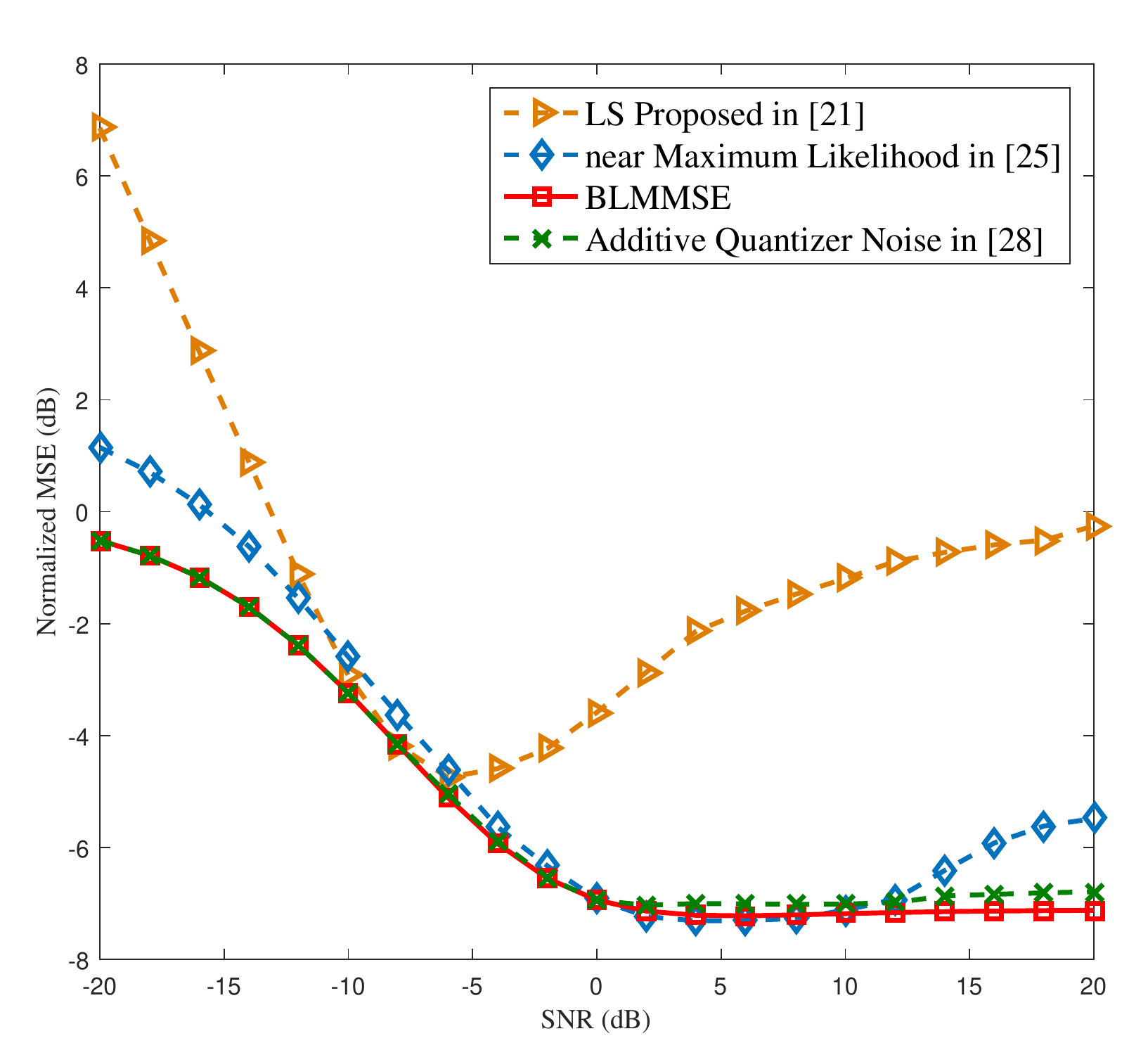}\\
  \vspace{-0.3cm}
  \caption{MSE of channel estimators versus SNR with $M=16, K = 4$ and $\tau = 20$. Least Squares estimator is from \cite{chiara2014massive} and nML estimator is from~\cite{juncil2015near}.}\label{CE_LS_nML_MMSE}
  \vspace{-0.5cm}
\end{figure}

Figure \ref{CE_LS_nML_MMSE} compares the MSE of the various channel estimators as a function of SNR for a case with $M=16$, $K= 4$ and $\tau = 20$. {\color{black} Note that we also include the performance of a similar channel estimator proposed in \cite{mollen2016performance}, in which the quantizer noise is modeled as uncorrelated additive noise with a covariance matrix $\mathbf{C}_{\mathbf{q}_{\rm p}} = (1-2/\pi){\color{black}\mathbf{I}_{M\tau}}$. We emphasize again that in our work, the correlation between the elements of the quantization noise vector is taken into account using the arcsine law, and hence $\mathbf{C}_{\mathbf{q}_{\rm p}}$ is not in general a diagonal matrix. We see that our proposed BLMMSE approach outperforms the other previously proposed approaches.  We also see that at low SNR, BLMMSE and the method based on uncorrelated quantization noise achieves the same performance, which verifies the observation that the approximation of \eqref{C_qq_app} is reasonable at low SNR. However, with the increase of SNR, a small performance gap can be seen between these two curves, indicating that not considering the correlation between the quantizer noise in one-bit systems may cause performance loss and the correlation should be taken into account.

{\color{black}A larger gap will result in cases where the quantizer noise is spatially correlated, since the analysis of \cite{mollen2016performance} did not take this possibility into account. This will occur for example if the channel or the additive noise is itself spatially correlated. For example, take the simple case depicted in Fig.~\ref{SpatiallyCorrelated} for $M=16, K=1$ and $\tau=2$, which shows the MSE performance for a case with a spatially correlated channel where $\mathbf{C}_{\underline{\mathbf{h}}}$ is non-diagonal. In this case, the BLMMSE channel estimator is given by
\begin{align}\label{BLMMSE_SpatiallyCorrelated}
\underline{\hat{\mathbf{h}}}^{\texttt{BLM}} = \mathbf{C}_{\underline{\mathbf{h}}}(\mathbf{A}_{\rm p}\bar{\bm{\Phi}})^H\mathbf{C}_{\mathbf{r}_{\rm p}}^{-1}\mathbf{r}_{\rm p},
\end{align}
where, following the same step as in \eqref{optimal_A_p}, the matrix $\mathbf{A}_{\rm p}$ is
\begin{align}
\mathbf{A}_{\rm p} = \sqrt{\frac{2}{\pi}}\diag\left(\bar{\bm{\Phi}}\mathbf{C}_{\underline{\mathbf{h}}} \bar{\bm{\Phi}}^H + {\color{black}\mathbf{I}_{M\tau}}\right)^{-\frac{1}{2}}.
\end{align}
For this example, we consider a typical urban channel model as described in \cite{klaus2000a}, where the power angle spectrum of the channel is modeled by a Laplacian distribution with an angle spread of $10^\circ$. The covariance matrix $\mathbf{C}_{\underline{\mathbf{h}}}$ can then be obtained according to \cite[Eq. (2)]{you2015pilot}.
We can see that the MSE performance gap grows to over 1~dB, indicating that the spatial correlation of the quantizer noise has an impact on performance and should be taken into account.}

\begin{figure}[!t]
  \centering
  \includegraphics[width=8cm]{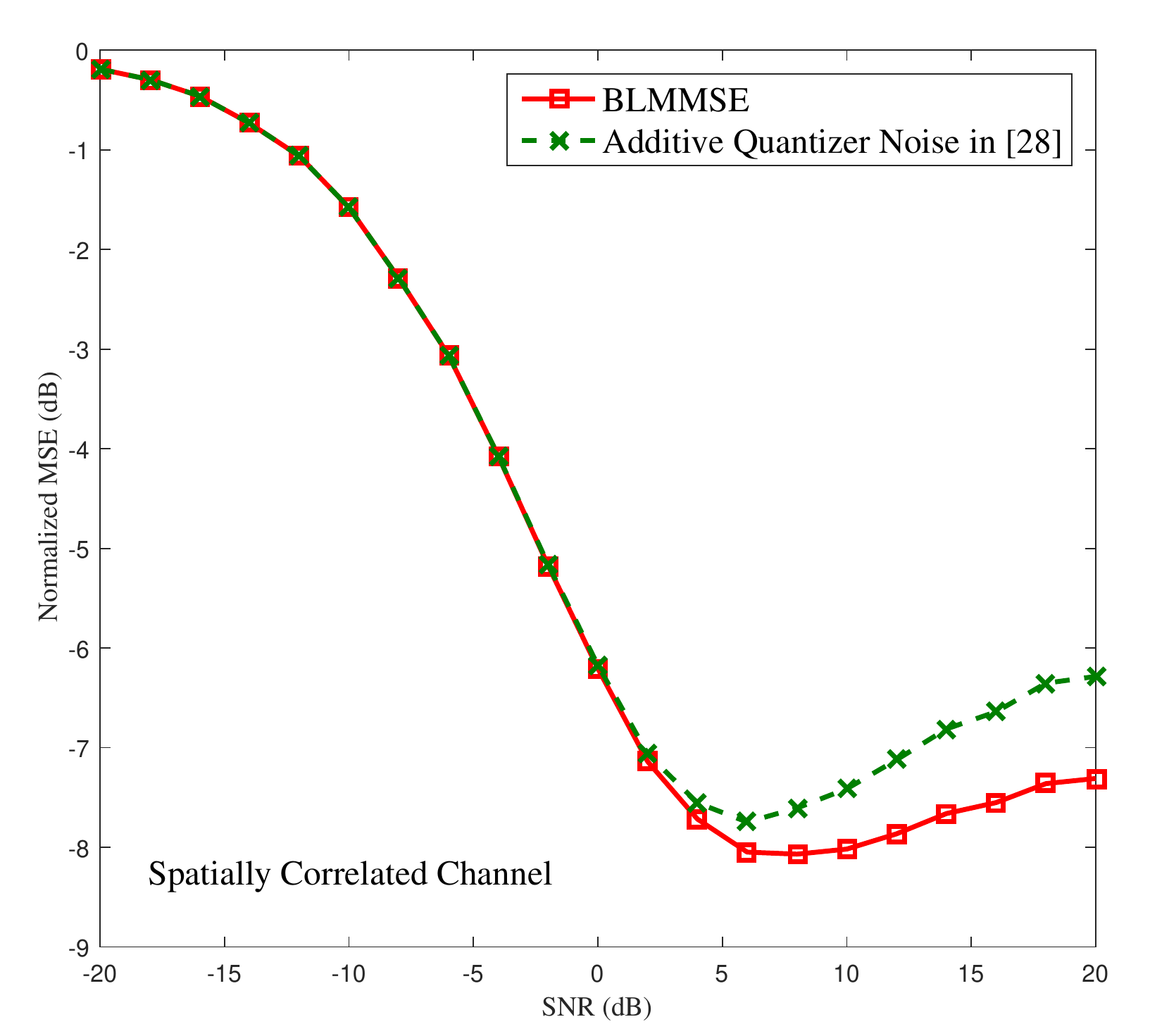}\\
  \vspace{-0.3cm}
  \caption{MSE of channel estimators versus SNR with $M=16, K=1$ and $\tau = 2$ over a spatially correlated channel. }\label{SpatiallyCorrelated}
  \vspace{-0.5cm}
\end{figure}

}

\subsection{Validation of Achievable Rate Results}
\begin{figure}[!t]
  \centering
  \includegraphics[width=8cm]{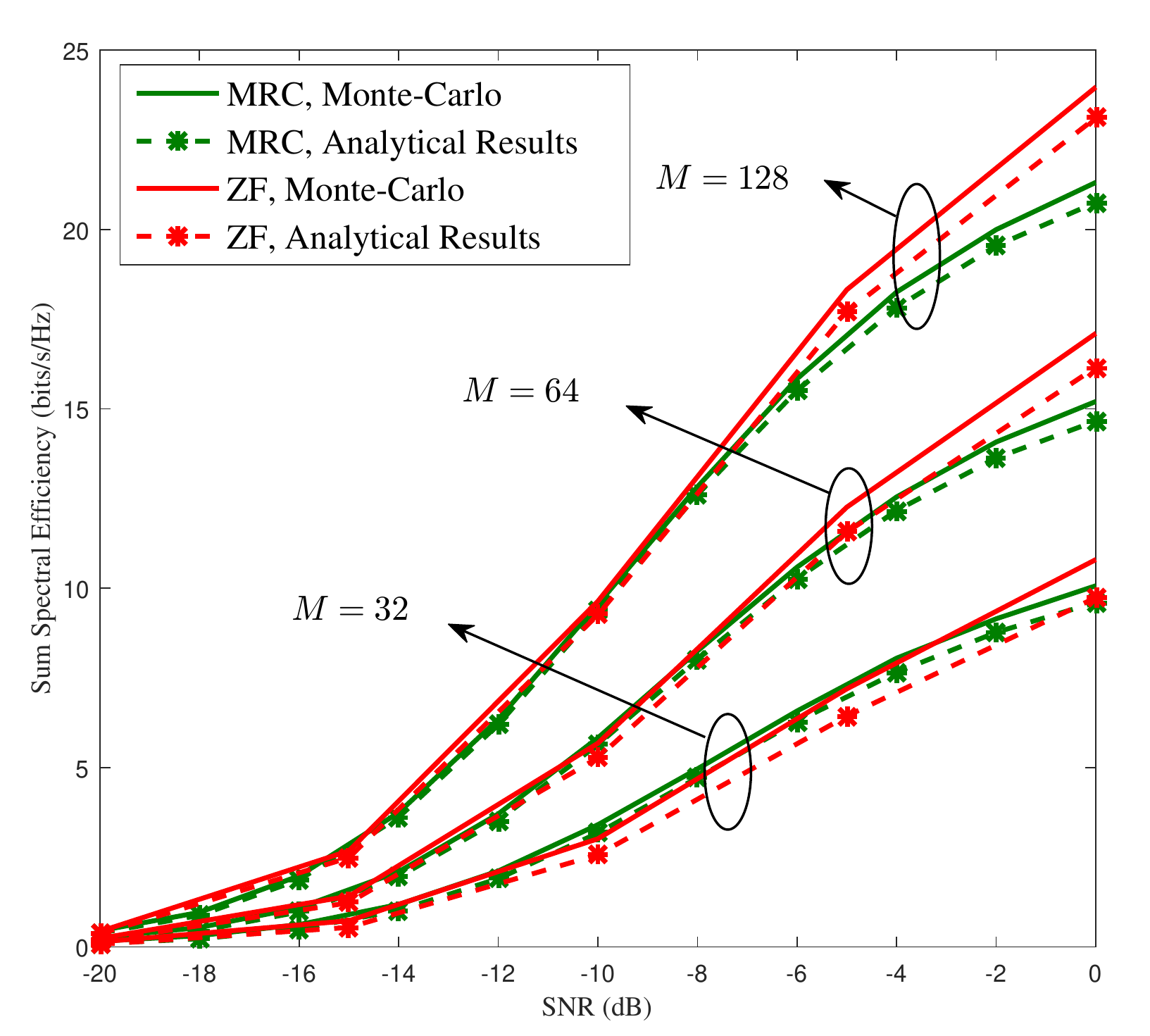}\\
  \vspace{-0.5cm}
  \caption{Sum spectral efficiency versus SNR with $M=\{32,64,128\}$ and $K =\tau= 8$ for MRC and ZF receivers.}\label{SE_MRC_ZF}
  \vspace{-0.3cm}
\end{figure}
{\color{black}Here we evaluate the validity of the lower bounds on the achievable rate for the MRC and ZF receivers derived in Theorems 1 and 2 compared with the ergodic rate given in \eqref{ergodic_achievable_rate}.} Fig.~\ref{SE_MRC_ZF} shows the sum spectral efficiency versus SNR with $K=\tau = 8$ for different numbers of transmit antennas $M=\{32, 64, 128\}$. {\color{black}The dashed lines represent the sum spectral efficiencies obtained using the closed-form expressions in~\eqref{MRC_Achievable_Rate} and~\eqref{ZF_Achievable_Rate} for the MRC and ZF receivers, respectively}, while the solid lines represent the ergodic sum spectral efficiencies obtained from \eqref{ergodic_achievable_rate}.
For both the MRC and ZF receiver, the gap between the approximation and the lower bound of the ergodic rate  is small. {\color{black}For example, with $M = 128$ and $\rm{SNR} = -10$dB, the sum spectral efficiency gap is 0.19 bits/s/Hz and 0.38 bits/s/Hz for the MRC and ZF receivers, respectively.} This implies that the approximation on the achievable rate given in \eqref{Achievable_Rate} is a good predictor of the performance of one-bit massive MIMO systems. Thus, in the following plots we will show only the approximation when evaluating performance.

\subsection{One-Bit Massive MIMO Power Efficiency}
\begin{figure}[!t]
  \centering
  \includegraphics[width=8cm]{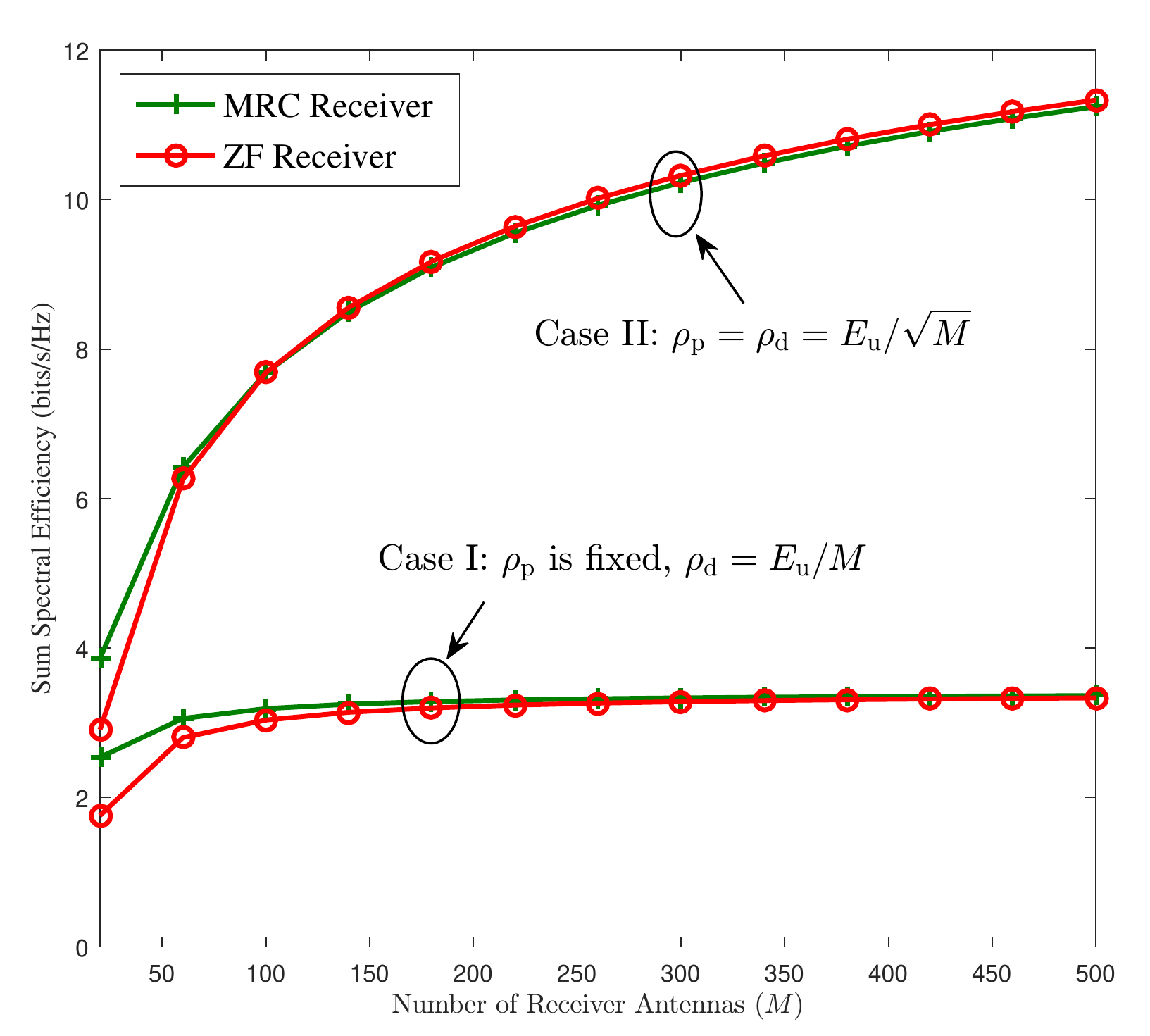}\\
  \vspace{-0.3cm}
  \caption{Sum spectral efficiency versus number of BS antennas $M$ for MRC and ZF receivers with $\rho_{\rm p} = 0$dB, $\rho_{\rm d} = E_{\rm u}/M$ in Case I, and $\rho_{\rm p} = \rho_{\rm d} = E_{\rm u}/\sqrt{M}$ in Case II.}\label{PowerEfficiency}
  \vspace{-0.5cm}
\end{figure}

This example considers the power efficiency of using large antenna arrays in one-bit massive MIMO for the two cases considered in Section~\ref{sec:power}. Fig.~\ref{PowerEfficiency} shows the sum spectral efficiency versus the number of receive antennas with $K=\tau = 8$ for the MRC and ZF receivers for Cases I and II. In Case I, we assume $\rho_{\rm p} = 10$dB is fixed and $\rho_{\rm d} = E_{\rm u}/M$, while in Case II we choose $\rho_{\rm p} = \rho_{\rm d} = E_{\rm u}/\sqrt{M}$, where $E_{\rm u} = 0$dB. As predicted by the analysis, in Case I the sum spectral efficiency converges to the same constant value for both the MRC and ZF receivers. In Case II where $\rho_{\rm p}= \rho_{\rm d} = E_{\rm u}/\sqrt{M}$, the sum spectral efficiency also converges to a constant value for both the MRC and ZF receivers, although the constant is only reached for very large $M$.

\subsection{Resource Allocation}

\begin{figure}[!t]
  \centering
  \includegraphics[width=8cm]{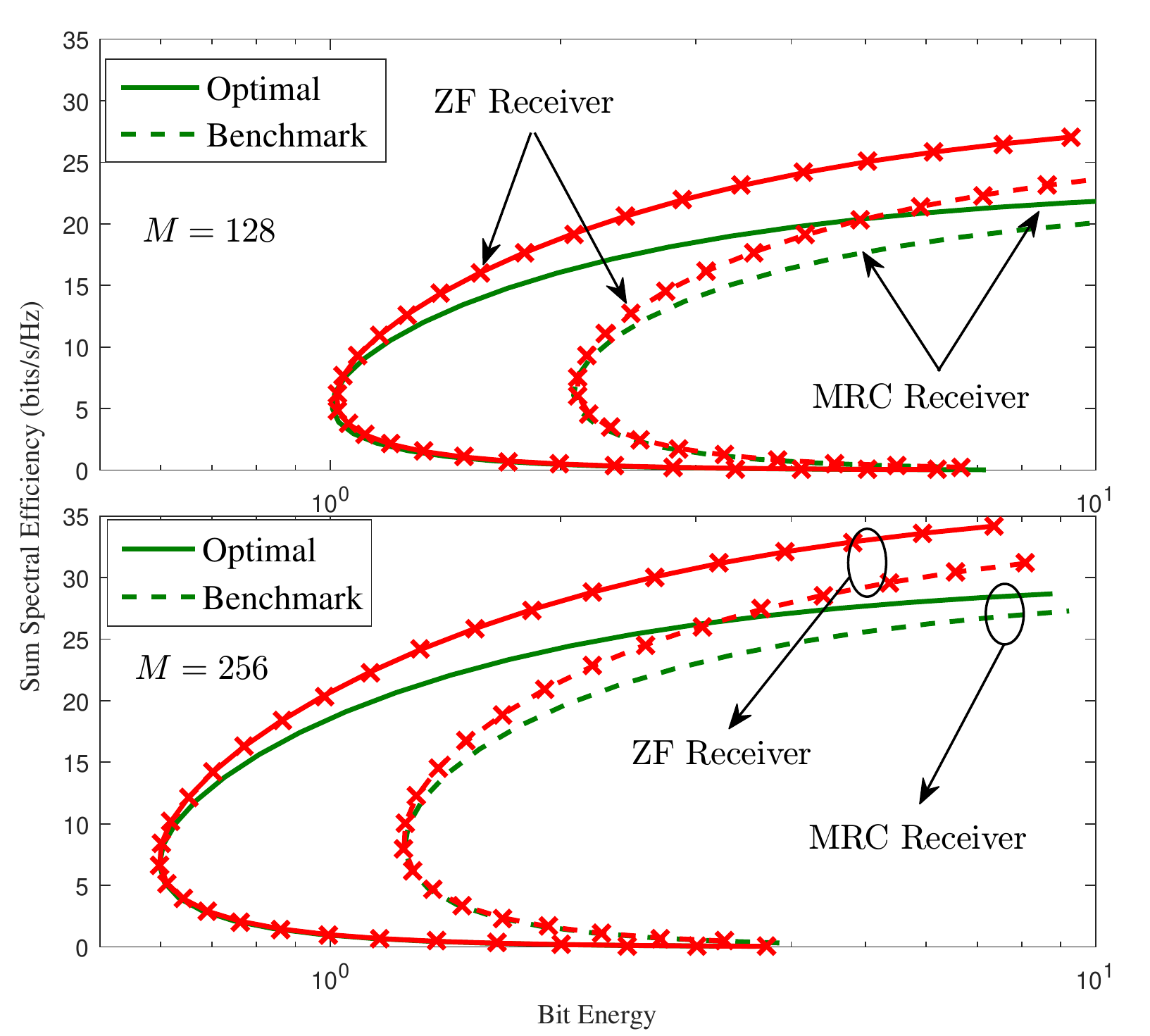}\\
  \vspace{-0.3cm}
  \caption{Bit energy versus sum spectral efficiency with and without resource allocation for $M=\{128, 256\}$ and $K=8$.}\label{PowerAllocation}
  \vspace{-0.3cm}
\end{figure}

{\color{black}We now investigate the benefit of our proposed optimal resource allocation scheme that adjusts the training length, training power, and data transmission power.}  In order to illustrate the benefit achieved by our proposed allocation scheme, we define the bit energy as the total transmit power expended divided by the sum spectral efficiency, or energy consumed per transmitted bit:
\begin{equation}\label{Bit_Energy}
\zeta_{\texttt{A}} = \frac{\tau\rho_{\rm p} + (T-\tau)\rho_{\rm d}}{\mathcal{S}_{\texttt{A}}}.
\end{equation}
Fig.~\ref{PowerAllocation} shows the sum spectral efficiency versus the bit energy with and without optimal power allocation for $M=\{128,256\}$ and for the MRC and ZF receivers. {\color{black}The `Benchmark' curves correspond to choosing $\tau = K$ and $\rho_{\rm p} = \rho_{\rm d}$, while the `Optimal' curves are obtained using the optimal resource allocation of \eqref{opt_BE2}.} Different points on the curves correspond to different values of total available power. The benefit of an optimal power allocation is very evident in all cases.  For example, to achieve a sum spectral efficiency of $15$ bits/s/Hz with $M=128$, {\color{black}the optimal resource allocation can reduce the bit energy by a factor of 1.9 for both the MRC and ZF receivers compared to the benchmark case.} The improvement in bit energy achieved by increasing the number of antennas is also apparent. For a sum spectral efficiency of $15$ bits/s/Hz and using the optimal resource allocation, {\color{black}we can reduce the bit energy by a factor of about 2.2 for the MRC and ZF receivers, by doubling the number of antennas from 128 to 256.}

\begin{figure}[!t]
  \centering
  \includegraphics[width=8cm]{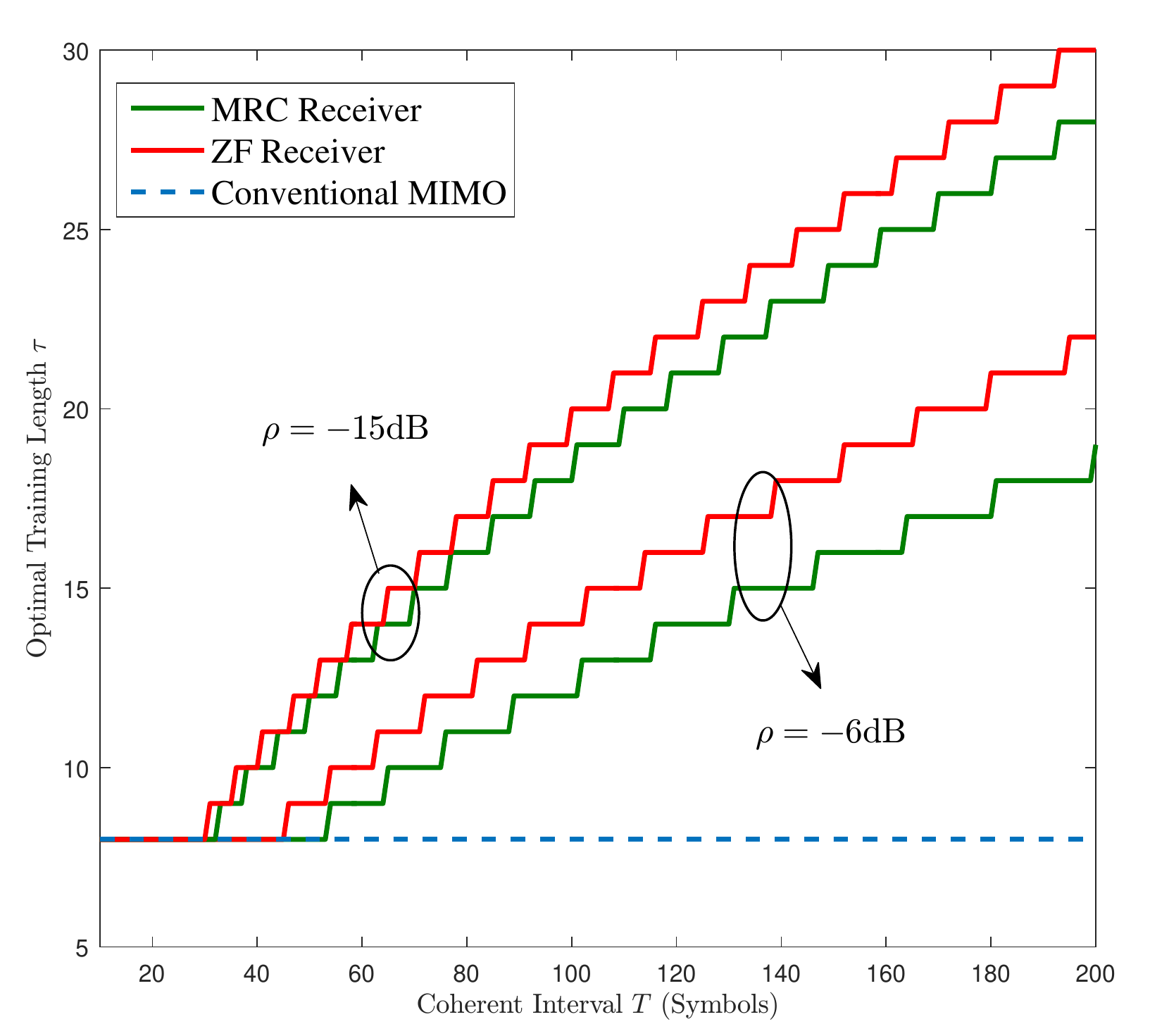}\\
  \vspace{-0.3cm}
  \caption{Optimal training length versus the coherence interval with $M=128$ and average transmit power $\rho=\{-15,-6\}$dB for the MRC and ZF receivers.}\label{OptimalTraining}
  \vspace{-0.3cm}
\end{figure}


{\color{black}Fig.~\ref{OptimalTraining} shows the optimal training duration versus the length of the coherence interval for $M = 128$, $K=8$ and average transmit power $\rho=\{-15,-6\}$dB for conventional and one-bit massive MIMO systems. We can see that the optimal training length is always equal to the number of users for conventional massive MIMO systems, while it depends on the coherence interval and the total power budget for one-bit MIMO systems. This is because a larger proportion of the coherence interval devoted to training is required in one-bit systems to combat the quantization noise. In addition, we observe that the optimal training length for the MRC receiver is smaller than that for the ZF receiver, implying that the ZF receiver demands a higher quality channel estimate than MRC in order to reduce the interuser interference, and hence improve the sum spectral efficiency.}

\subsection{Number of Antennas for One-Bit and Conventional Massive MIMO}
\begin{figure}[!t]
  \centering
  \includegraphics[width=8cm]{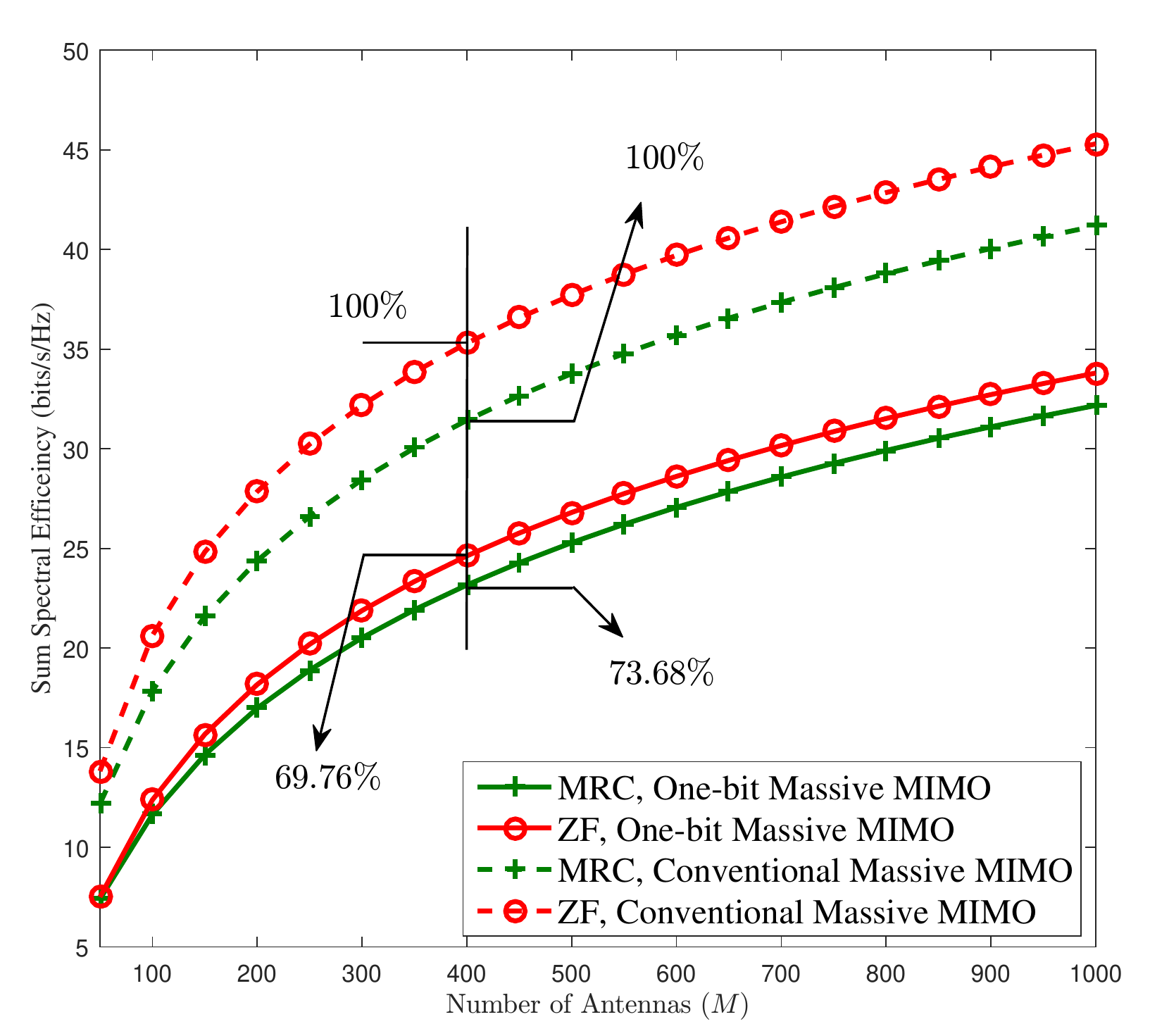}\\
  \vspace{-0.3cm}
  \caption{Comparison of the sum spectral efficiency versus number of receive antennas for one-bit and conventional massive MIMO systems with average transmit power $\rho= -10$dB.}\label{Compare_One_Typ_SE_vs_M}
  \vspace{-0.5cm}
\end{figure}
In this example we compare the sum spectral efficiencies between one-bit and conventional massive MIMO systems. Fig.~\ref{Compare_One_Typ_SE_vs_M} illustrates the sum spectral efficiency versus the number of receive antennas for the MRC and ZF receivers with an average transmit power $\rho=-10$dB. {\color{black}Since we are more interested in comparing the maximum sum spectral efficiencies of both one-bit and conventional systems, each curve is obtained by adjusting the training length, the training power and data transmission power to maximize the sum spectral efficiency, as in problem \eqref{opt_BE2}.} The curves for `Conventional massive MIMO' are obtained using the formulas in Table \ref{Table_C}. Compared with the conventional system, the rate loss of the one-bit system is not as severe as might be imagined. {\color{black}For example, with $M=400$, the one-bit system can still achieve a sum spectral efficiency of 23.2 bits/s/Hz and 24.6 bits/s/Hz for the MRC and ZF receivers, respectively, which amounts to $73.68\%$ and $69.76\%$ of the sum spectral efficiency of the conventional system. }This is a remarkably high value for such a coarsely quantized signal that only retains sign information about the received signals. The figure also verifies the increase in the number of antennas required for the one-bit system with MRC to achieve performance equivalent to a conventional massive MIMO system; {\color{black}the one-bit system requires about 480 antennas, or approximately $480/215 = 2.23$ times more antennas than for a conventional system to achieve a spectral efficiency of 25 bits/s/Hz.}

\begin{figure}[!t]
  \centering
  \includegraphics[width=8cm]{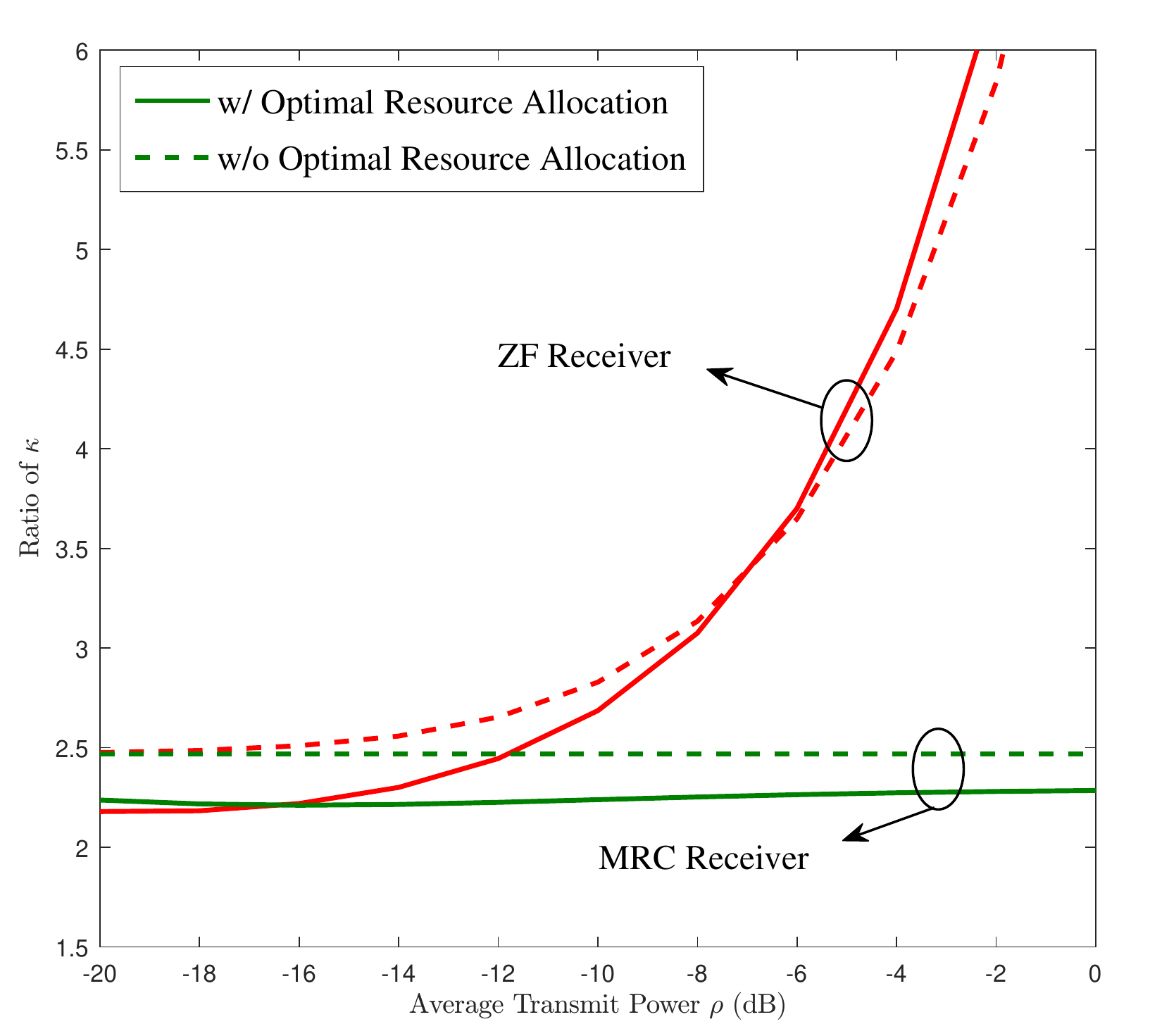}\\
  \vspace{-0.3cm}
  \caption{The ratio of $\kappa$ versus the average transmit power $\rho$ with $K=8$ for the MRC and ZF receivers.}\label{Compare_One_Typ_SE_Ratio}
  \vspace{-0.3cm}
\end{figure}

This relationship is further illustrated in~Fig. \ref{Compare_One_Typ_SE_Ratio} which shows the ratio of $\kappa = M_{\rm one}/M_{\rm conv}$ needed for the two types of systems to achieve equivalent performance. {\color{black}The curves labeled `w/o Optimal Resource Allocation' are obtained assuming $\tau = K$ and $\rho_{\rm p}=\rho_{\rm d}=\rho$, while the curves labeled `w/ Optimal Resource Allocation' are obtained by solving problem \eqref{opt_Ratio}. We can see that the ratio is constant at 2.5 for MRC and also at low SNR for ZF for the case without resource allocation, which verifies the conclusion in \cite{mollen2016performance}.  However, for the case with an optimal resource allocation, the ratio is around 2.2-2.3, which implies that fewer antennas are needed for the one-bit system if its performance is optimized. In addition, we see that as the average transmit power $\rho$ increases, the number of antennas required for a one-bit system to have equivalent performance with the ZF receiver grows without bound,} since the conventional ZF receiver is theoretically able to obtain a better and better channel estimate that allows it to ultimately eliminate all inter-user interference.

\section{Conclusions}
This paper has investigated channel estimation and overall system performance for the single-cell, flat Rayleigh fading massive MIMO uplink when one-bit ADCs are employed at the BS. We  used the Bussgang decomposition to derive a new channel estimator based on the LMMSE criteria, and showed that the resulting BLMMSE estimator provides the lowest MSE among various competing algorithms. However, even the BLMMSE estimator has a high-SNR error floor due to the one-bit quantization. We derived simple closed-form approximations for {\color{black}the massive MIMO uplink achievable rate for low SNR and a large number of users} assuming MRC and ZF receivers that employ the BLMMSE channel estimate. We then used the approximation to study the sum spectral efficiency and energy efficiency of the one-bit massive MIMO uplink. Our results show that massive MIMO still yields similar gains in energy efficiency when one-bit quantizers are employed, and we developed an optimization problem that when solved yields significant gains in spectral efficiency by properly selecting the training length, training power and data transmission power. {\color{black}We showed that for an MRC receiver with optimal resource allocation, approximately 2.2-2.3 times more antennas are required in a one-bit massive MIMO system to achieve the same spectral efficiency as a conventional system with full-precision ADCs.} However, significantly more antennas are required in a one-bit system for the ZF receiver at high SNR. Finally, we presented a number of simulation results that validate our analysis and illustrate the potential performance of massive MIMO systems with one-bit ADCs.

\appendices
\section{}\label{proof_uncorrelation}
For a given $\mathbf{y}_{\rm p}$, the covariance matrix between the quantizer noise $\mathbf{q}_{\rm p}$ and the channel vector {\color{black}${\underline{\mathbf{h}}}$} can be expressed as
\begin{align}\label{E_qh}
\E\left\{\mathbf{q}_{\rm p}{\color{black}\underline{\mathbf{h}}^H}\right\} = \E_{\mathbf{y}_{\rm p}}\left\{\E\left\{\mathbf{q}_{\rm p}{\color{black}\underline{\mathbf{h}}^H}|\mathbf{y}_{\rm p}\right\}\right\}  .
\end{align}
Since the quantizer noise $\mathbf{q}_{\rm p} = \mathbf{r}_{\rm p} - \mathbf{A}_{\rm p}\mathbf{y}_{\rm p}$ is fixed for a given $\mathbf{y}_{\rm p}$, we can remove $\mathbf{q}_{\rm p}$ from the inner expectation of~\eqref{E_qh} to obtain
\begin{align}
\E_{\mathbf{y}_{\rm p}}\left\{\E\left\{\mathbf{q}_{\rm p}{\color{black}\underline{\mathbf{h}}^H}|\mathbf{y}_{\rm p}\right\}\right\} = \E_{\mathbf{y}_{\rm p}}\left\{\mathbf{q}_{\rm p}\E\left\{{\color{black}\underline{\mathbf{h}}^H}|\mathbf{y}_{\rm p}\right\}\right\}.
\end{align}
According to \cite{kay1993fundamentals}, the value of $\E\left\{{\color{black}\underline{\mathbf{h}}^H}|\mathbf{y}_{\rm p}\right\}$ is the linear MMSE estimate of ${\color{black}\underline{\mathbf{h}}}$, leading to
\begin{align}
\E_{\mathbf{y}_{\rm p}}\left\{\E\left\{\mathbf{q}_{\rm p}{\color{black}\underline{\mathbf{h}}^H}|\mathbf{y}_{\rm p}\right\}\right\} = \E_{\mathbf{y}_{\rm p}}\left\{\mathbf{q}_{\rm p}\mathbf{y}_{\rm p}^H\mathbf{C}_{\mathbf{y}_{\rm p}}^{-1}\mathbf{C}_{\mathbf{y}_{\rm p}{\color{black}\underline{\mathbf{h}}}}\right\}.
\end{align}
Choosing $\mathbf{A}_{\rm p}$ according to \eqref{optimal_A_p}, the quantizer noise $\mathbf{q}_{\rm p}$ is uncorrelated with $\mathbf{y}_{\rm p}$, and hence we have
\begin{align}
\E\left\{\mathbf{q}_{\rm p}{\color{black}\underline{\mathbf{h}}^H}\right\} &=\E_{\mathbf{y}_{\rm p}}\left\{\mathbf{q}_{\rm p}\mathbf{y}_{\rm p}^H\mathbf{C}_{\mathbf{y}_{\rm p}}^{-1}\mathbf{C}_{\mathbf{y}_{\rm p}{\color{black}\underline{\mathbf{h}}}}\right\} = \mathbf{0},
\end{align}
which implies that the quantizer noise $\mathbf{q}_{\rm p}$ is uncorrelated with the channel $\underline{\mathbf{h}}$.

\section{}\label{proof_Lemma1}
We follow the approach of \cite{medard2000the} and only exploit knowledge of the average effective channel ${\rm{E}}\left\{ {\sqrt {{\rho _{\rm d}}} \mathbf{w}_k^T{{\bf{A}}_{\rm d}}{{\bf{h}}_k}} \right\}$ in the detection. Then, according to \cite{hassibi2003how}, the lower bound of the achievable rate in \eqref{Achievable_Rate} is obtained by treating the uncorrelated inter-user interference and the quantizer noise as independent Gaussian noise, which is a worst-case assumption when computing the mutual information \cite{hassibi2003how}. Therefore the variance of the effective noise is
\begin{align}\label{variance_effective_noise}
\E\{|\tilde{n}_{{\rm d},k}|^2\} =& \Var\left\{\left({\mathbf{w}_k^T  \mathbf{A}_{\rm d}{{\bf{h}}_k}}\right)\right\} + \rho_{\rm d}\sum_{i\neq k}^K{\E\left\{\left|{\mathbf{w}_k^T \mathbf{A}_{\rm d}{{\bf{h}}_i}}\right|^2\right\}} \nonumber\\
& +\E\left\{\left\|\mathbf{w}_k^T \mathbf{A}_{\rm d} \right\|^2\right\} +\E\left\{\mathbf{w}_k^T \mathbf{C}_{\mathbf{q}_{\rm d}}\mathbf{w}_k^T\right\},
\end{align}
where the expectation operation is taken with respect to the channel realizations.
{\color{black}By using the same result in \eqref{C_qq_app} at low SNR, we can approximate the quantizer noise as
\begin{align}\label{quantizer_noise_receive}
\E\left\{\mathbf{w}_k^T \mathbf{C}_{\mathbf{q}_{\rm d}}\mathbf{w}_k^T\right\} =  \left(1-\frac{2}{\pi}\right)\E\left\{\left\|\mathbf{w}_k^T\right\|^2\right\}.
\end{align}
}
Substituting $\mathbf{A}_{\rm d} = \alpha_{\rm d}{\color{black}\mathbf{I}_{K}}$ and combining \eqref{variance_effective_noise} and \eqref{quantizer_noise_receive}, we arrive at Lemma 1.

\section{}\label{proof_theorem2}
From \eqref{Achievable_Rate}, we need to compute ${\E\left\{ {\mathbf{w}_k^T {\mathbf{h}_k}} \right\}}$, $\Var\left({\mathbf{w}_k^T {{\bf{h}}_k}}\right)$, ${\rm{UI}}_k$ {\color{black}and ${\rm{AQN}}_k$}. Note that, although the channel vector $\mathbf{h}_k$ is Gaussian, the BLMMSE channel estimate $\hat{{\mathbf{h}}}_k$ is not Gaussian due to the quantizer noise. However, we can approximate $\hat{{\mathbf{h}}}_k$ as  Gaussian using Cram\'{e}r's central limit theorem \cite{Cramer2004random}.

For the MRC receiver $\mathbf{W}^T_{\texttt{MRC}} = \hat{{\mathbf{H}}}^H$, we have
\begin{equation}
      \mathbf{w}^T_k \mathbf{h}_k = \hat{{\mathbf{h}}}_k^H \mathbf{h}_k = \left\|\hat{{\mathbf{h}}}_k\right\|^2 + \hat{{\mathbf{h}}}_k^H \bm{\varepsilon}_{k}.
\end{equation}
Therefore,
\begin{equation}\label{MRC_Proof_denominator}
       \E\left\{\hat{{\mathbf{h}}}_k^H \mathbf{h}_k\right\} = \E\left\{\left\|\hat{{\mathbf{h}}}_k\right\|^2\right\} = M\sigma^2.
\end{equation}
The variance of ${\mathbf{w}_k^T {{\bf{h}}_k}}$ is given by
\begin{align}
  \Var&\left({\mathbf{w}_k^T {{\bf{h}}_k}}\right)= \E\left\{\left\|\hat{{\mathbf{h}}}_k^H \mathbf{h}_k\right\|^2\right\} - M^2 \sigma^4 \nonumber \\
  & = \E\left\{\left\|\hat{{\mathbf{h}}}_k\right\|^4\right\} + \E\left\{\left\|\hat{{\mathbf{h}}}_k^H{\bm{\varepsilon}_k}\right\|^2\right\} - M^2 \sigma^4.
\end{align}

Since $\hat{{\mathbf{h}}}_k$ is approximately Gaussian with variance of each element $M\sigma^2$, we obtain
\begin{align}\label{MRC_Proof_var}
  \Var&\left({\mathbf{w}_k^T {{\bf{h}}_k}}\right)= \sigma^4 M(M+1) + \sigma^2 (1-\sigma^2)M - M^2\sigma^4 \nonumber \\
  & = M\sigma^2.
\end{align}
For $i\neq k $ we have
\begin{align}\label{MRC_Proof_UI}
  {\rm{UI}}_k = \rho_{\rm d}\alpha_{\rm d}^2\sum_{i\neq k }^K \E\left\{\left| \hat{{\mathbf{h}}}_k^H \mathbf{h}_i\right|^2\right\} = (K-1)\rho_{\rm d}\alpha_{\rm d}^2M\sigma^2.
\end{align}
Similarly, we obtain
\begin{align}\label{MRC_Proof_ANQN}
  {\rm{AQN}}_k &= (\alpha_{\rm d}^2+1-2/\pi)M\sigma^2.
\end{align}
Substituting \eqref{MRC_Proof_denominator}, \eqref{MRC_Proof_var}, \eqref{MRC_Proof_UI} and \eqref{MRC_Proof_ANQN} into \eqref{Achievable_Rate}, {Theorem 1} is obtained.

\section{}\label{proof_theorem3}
For the ZF receiver $\mathbf{W}^T_{\texttt{ZF}} =\left({\hat{\mathbf{H}}}^H{\hat{\mathbf{H}}}\right)^{-1}{\hat{\mathbf{H}}}^H$, we have
\begin{equation}
\mathbf{W}^T_{\texttt{ZF}}\mathbf{H} = \mathbf{W}^T_{\texttt{ZF}}(\hat{{\mathbf{H}}} + \mathbf{\mathcal{E}}) = {\color{black}\mathbf{I}_{K}} + \mathbf{W}^T_{\texttt{ZF}}\mathbf{\mathcal{E}}.
\end{equation}
Therefore,
\begin{equation}\label{ZF_Proof_basic}
\mathbf{w}^T_{\texttt{ZF},k}\mathbf{h}_k = 1+\mathbf{w}^T_{\texttt{ZF},k}\bm{\varepsilon}_k.
\end{equation}
Similar to the derivation of the MRC receiver, we need to compute ${\E\left\{ {\mathbf{w}_k^T {\mathbf{h}_k}} \right\}}$, $\Var\left({\mathbf{w}_k^T {{\bf{h}}_k}}\right)$, ${\rm UI}_k$ and {\color{black}${\rm AQN}_k$}.

For the ZF receiver, we have
\begin{equation}\label{ZF_Proof_denominator}
       \E\left\{ {\mathbf{w}_k^T {\mathbf{h}_k}} \right\} = 1+\E\left\{\mathbf{w}^T_{\texttt{ZF},k}\bm{\varepsilon}_k\right\} = 1.
\end{equation}
The variance of ${\mathbf{w}_k^T {{\bf{h}}_k}}$ is given by
\begin{align}
  \Var&\left({\mathbf{w}_k^T {{\bf{h}}_k}}\right)= \E\left\{\left\|\mathbf{w}^T_{\texttt{ZF},k}\bm{\varepsilon}_k\right\|^2\right\} \nonumber \\
  & = (1-\sigma^2)\E\left\{\left\|\mathbf{w}^T_{\texttt{ZF},k}\right\|^2\right\} \nonumber \\
  & = (1-\sigma^2)\E\left\{\left[\left(\hat{{\mathbf{H}}}^H\hat{{\mathbf{H}}}\right)^{-1}\right]_{k,k}\right\}.
\end{align}
Since $\hat{{\mathbf{H}}}$ is approximately Gaussian, $\hat{{\mathbf{H}}}^H\hat{{\mathbf{H}}}$ is a $K\times K$ central Wishart matrix with $M$ degrees of freedom, Thus,
\begin{align}\label{ZF_Proof_var}
  \Var&\left({\mathbf{w}_k^T {{\bf{h}}_k}}\right)= \frac{(1-\sigma^2)}{\sigma^2(M-K)}.
\end{align}

From \eqref{ZF_Proof_basic}, for $i\neq k $ we have
\begin{align}\label{ZF_Proof_UI}
  {\rm{UI}}_k &= \rho_{\rm d}\alpha_{\rm d}^2\sum_{i\neq k }^K \E\left\{\left| \hat{{\mathbf{h}}}_k^H \bm{\varepsilon}_i\right|^2\right\} \nonumber \\
  &= \rho_{\rm d}\alpha_{\rm d}^2\sum_{i\neq k }^K(1-\sigma^2)\E\left\{\left[\left(\hat{{\mathbf{H}}}^H\hat{{\mathbf{H}}}\right)^{-1}\right]_{k,k}\right\} \nonumber \\
  & =(K-1)\rho_{\rm d}\alpha_{\rm d}^2\frac{(1-\sigma^2)}{\sigma^2(M-K)}.
\end{align}
Similarly,
\begin{align}\label{ZF_Proof_ANQN}
  {\rm{AQN}}_k &=\frac{\alpha_{\rm d}^2+1-2/\pi}{\sigma^2(M-K)}.
\end{align}
Substituting \eqref{ZF_Proof_denominator}, \eqref{ZF_Proof_var}, \eqref{ZF_Proof_UI} and \eqref{ZF_Proof_ANQN} into \eqref{Achievable_Rate}, {Theorem 2} is obtained.

\section{}\label{proof_Lemma2}
{\color{black}First we rewrite the sum spectral efficiency of \eqref{MRC_Achievable_Rate} and \eqref{ZF_Achievable_Rate} for the MRC and ZF receivers as a function with respect to $\gamma$ and $\tau$:
\begin{equation}\label{eq_lemma2}
\mathcal{S}_{\texttt A}(\gamma,\tau) = \frac{T-\tau}{T} K \log_2\left(1+\frac{a_1 \tau}{a_2 \tau^2 +a_3 \tau + a_4}\right),
\end{equation}
where we define
\begin{align}
a_1 &= 4 M P^2 (\gamma - \gamma^2) ~~~~~~~a_2 = \pi^2 + 2 \pi P\gamma \nonumber \\
a_3 &= \pi(K P (\pi-2) \gamma - K P (1 - \gamma) (\pi + 2 P \gamma) - (\pi + 2 P \gamma) T) \nonumber \\
a_4 &= \pi (K^2 P^2 (\pi-2) (-1 + \gamma)\gamma  - K P (\pi-2) \gamma T) \nonumber
\end{align}
for ${\texttt A} = {\texttt {MRC}}$, and
\begin{align}
a_1 &= 4 (M-K) P^2 (\gamma - \gamma^2) ~~~~~~~a_2 = \pi^2 + 2\pi P \gamma \nonumber \\
a_3 &= -K P (2 \pi (\gamma + P (\gamma - \gamma^2))  +
    4 P (\gamma - \gamma^2)+ \pi^2 (2 \gamma -1 )) \nonumber \\
    & -  (\pi^2 + 2 \pi P \gamma) T \nonumber \\
a_4 &= \pi (K^2 P^2 (\pi-2) (-1 + \gamma)\gamma  - K P (\pi-2) \gamma T) \nonumber
\end{align}
for ${\texttt A} = {\texttt {ZF}}$.

Then we denote $\{\gamma^*, \tau^*\}$ to be the solution of \eqref{opt_BE2}, such that $\gamma^* P = \tau^* \rho_{\rm p}^*$ is the optimal power for training, and $(1 - \gamma^*)P = (T - \tau^*)\rho_{\rm d}^*$ is the optimal amount for data transmission. Next we choose $\bar{\tau} = K, \bar{\rho}_p = \gamma^* P/\bar{\tau}$ and $\bar{\rho}_d = (1-\gamma^*)P/(T -\bar{\tau})$. Clearly, the function in \eqref{eq_lemma2} is not a monotonic function with respect to $\tau$ with a given $\gamma^*$. That is to say, it is difficult to compare the values of $\mathcal{S}(\gamma^*, \tau^*)$ and $\mathcal{S}(\gamma^*, \bar{\tau})$. Therefore, we conclude that the optimal training length is not always equal to the number of users for one-bit systems.
}

{\color{black}
\bibliographystyle{IEEEtran}
\bibliography{reference}
}
\end{document}